\documentclass[fleqn,usenatbib,useAMS]{mnras}

\usepackage{newtxtext,newtxmath}
\usepackage{mathrsfs}
\usepackage[utf8]{inputenc}
\usepackage{graphicx}                           
\usepackage{bm}
\usepackage{multirow}
\usepackage{multicol}        

\usepackage{amsmath}	
\usepackage{pdflscape}	
\usepackage[T1]{fontenc}
\usepackage{ae,aecompl}
\usepackage{ulem}
\usepackage[dvipsnames]{xcolor}
\usepackage{color}

\DeclareRobustCommand{\VAN}[3]{#2}
\let\VANthebibliography\thebibliography
\def\thebibliography{\DeclareRobustCommand{\VAN}[3]{##3}\VANthebibliography}

\title[]{Cosmological imprints in the filament with DisPerSE}

\title[Ziyong Wu et al.]{Cosmological imprints in the filament with DisPerSE}

\author[Ziyong Wu et al.]{
Ziyong Wu$^{1,2}$,
Yu Luo$^{3,1}$\thanks{luoyupmo@gmail.com},
Wei Wang$^{1,2}$,
Xi Kang$^{4,5,1}$\thanks{kangxi@zju.edu.cn},
Renyue Cen$^{5,4}$
\\
$^{1}$Purple Mountain Observatory, Chinese Academy of Sciences, 10 Yuanhua Road, Nanjing 210033, China\\
$^{2}$School of Astronomy and Space Sciences, University of Science and Technology of China, Hefei 230026, China\\
$^{3}$Department of Physics, School of Physics and Electronics, Hunan Normal University, Changsha 410081, China\\
$^{4}$Institute for Astronomy, The school of Physics, Zhejiang University, Hangzhou 310037, China\\
$^{5}$Center for Cosmology and Computational Astrophysics, Zhejiang University, Hangzhou 310027, China
}

\date{Accepted XXX. Received YYY; in original form ZZZ}

\pubyear{2015}

\begin{document}
\label{firstpage}
\pagerange{\pageref{firstpage}--\pageref{lastpage}}
\maketitle

\begin{abstract}
In the regime of cosmology and large-scale structure formation, filaments are vital components of the cosmic web. This study employs statistical methods to examine the formation, evolution, and cosmological constraints of filaments identified by DisPerSe. We run large-sample of N-body simulations to study the filament length and its evolution. In general, the filament length distribution can be fitted by a power law with both the normalization and power index dependent on redshift and cosmological parameters.  It is discovered that filament length distribution is influenced by various cosmological parameters, with $\sigma_8$ and $n_s$ exhibiting slightly stronger dependence than $\Omega_m$. We also uncover a three-stage filament formation process from $z \sim 3$ to $z \sim 1$: rapid formation of both long and short filaments from $z \sim 3$ to $z \sim 2$, persistence of long filaments from $z \sim 2$ to $z \sim 1$, followed by fragmentation and increased prevalence of shorter filaments below $z \sim 1$. Finally, we employ initial power spectrum fluctuations to elucidate the cosmological dependence on the filament length function. These insights enhance our understanding of filament evolution and their cosmological relevance and also highlight the potential cosmological applications in observations.
\end{abstract}




\section{Introduction}
\label{sec:Introduction}

The structure of the Universe on a large scale is in a web-like pattern referred to as the cosmic web \citep{1996Natur.380..603B} composed of clusters, filaments, walls, and voids. It is a colossal system of dark matter and baryon that emerges due to gravity's influence from the anisotropic collapse of initial density field fluctuations\citep{1970A&A.....5...84Z}. One of the principal challenges in observational and theoretical cosmology is to comprehend the physical properties and evolution of the different structures of the cosmic web.

Numerical simulations demonstrate that the cosmic web is characterized by the formation of nodes in dense regions, interconnected by filaments and sheets, while most regions are dominated by voids with incredibly low density. Filament plays an import role in large scale structures. Although galaxy clusters are relatively easier to detect and characterize within the context of large-scale structure and galaxy formation, it is predicted that filamentary structures contain the largest fraction of baryons. \citep{2010MNRAS.408.2163A,2006ApJ...650..560C,2014MNRAS.441.2923C}.
With the advent of wide-area spectroscopic redshift surveys, filaments has been detected by surveys such as the Sloan Digital Sky Survey (SDSS, \citet{2006AJ....131.2332G}), the Galaxy And Mass Assembly survey (GAMA, \citet{2011MNRAS.413..971D}), the Vimos Public Extragalactic Redshift Survey (VIPERS, \citet{2018A&A...609A..84S}),  two micron all sky survey (2MASS, \citet{2006AJ....131.2332G}) , and the COSMOS survey \citep{2018MNRAS.474.5437L}.
Therefore, filaments could serve as a complementary probe to galaxy clustering for cosmological constraints, underscoring the importance of studying their properties, evolution, and dependence on cosmological parameters and initial conditions.

Considerable efforts were made towards developing methods to detect and identify the filamentary structures. \citet{2018MNRAS.473.1195L} analysed the filaments detected by various methods and classified them into six groups based on their main characteristics:  graph and percolation techniques (e.g. MST, \citet{2014MNRAS.438..177A}, T-Rex method, \citet{2020A&A...637A..18B}), stochastic methods (e.g. BISOUS model, \citet{2016A&C....16...17T}), geometric Hessian-based methods (e.g. MMF, \citet{2007A&A...474..315A}, SHMAFF, \citet{2010MNRAS.409..156B}) , scalespace Hessian-based methods (e.g. NEXUS, \citet{2013MNRAS.429.1286C}), topological methods (e.g. DisPerSE, \citet{2011MNRAS.414..350S}, SpineWeb, \citet{2010ApJ...723..364A}), phase-space methods (e.g. ORIGAMI, \citet{2012ApJ...754..126F}). Meanwhile, as filaments are identified as the bridge of clusters by transferring matter to clusters \citep{2009MNRAS.396..635P}, it is also interesting to investigate the statistical properties of galaxies in the vicinity of cosmic filaments as well as exploring the interplay between filamentary environments and the properties of galaxies within them using different surveys such as GAMA \citep{2018MNRAS.474..547K}, COSMOS \citep{2018MNRAS.474.5437L}, VIPERS \citep{2017MNRAS.465.3817M}, SDSS \citep{2017MNRAS.466.1880C,2017A&A...600L...6K,2019MNRAS.483..223T,2019A&A...624A..48D}, which have found significant separations of galaxy masses and types, with massive or quench galaxies are closer to filament galaxies than smaller or active ones, underlining the fact that filaments play a role in shaping galaxy properties. Meanwhile, simulations such as HORIZON-AGN \citep{2019MNRAS.483.3227K}, ENZO \citep{2019MNRAS.486..981G}, 
IllustrisTNG \citep{2019MNRAS.486.3766M,2020A&A...641A.173G,2021A&A...649A.117G,2022A&A...661A.115G,2023A&A...671A.160G}, MillenniumTNG \citep{2024A&A...684A..63G}
also have been applied to study cosmic filaments. The implication from these studies is that the evolution of large-scale structures over cosmic time affects the potential wells of dark matter, ultimately influencing the formation of host galaxies. As a result, the infomration contained in filaments can serve as a valuable tool for measuring large-scale structures and testing cosmological theories.  \citet{2021MNRAS.507.2968W} utilizes the persistent points belonging to filaments to investigate the density range of filaments. Meanwhile, other researchers concentrate on analyzing the length, width, and shape of filaments, as well as their connections. Consistent analysis of observational data has revealed a wide range of filament lengths, ranging from short structures with lengths of approximately $5 h^{-1}\rm Mpc$ to exceedingly long objects that extend beyond $100 h^{-1}\rm Mpc$ \citep{2004ApJ...606...25B,2004MNRAS.354L..61P,2011MNRAS.411..332P,2014MNRAS.438.3465T}. \citet{2012MNRAS.422...25S} and \citet{2004MNRAS.354L..61P} have also observed that short filaments tend to be straight, while very extended objects, although they are prominent in the distribution of haloes and galaxies, are rare and represent only a small fraction of the filament population \citep{2010MNRAS.407.1449G,2014MNRAS.438.3465T}. \citet{2020A&A...641A.173G} discovered that filament density varies with filament length, with longer filaments being less dense consistently across various simulations. \citet{2010MNRAS.406..320C}, on the other hand, found that the distribution of dark matter filament width evolves from $z \sim 3$ to $z \sim 0$, with a distribution broadening and peaking at smaller widths as the universe expands. Meanwhile, the research of \citet{2014MNRAS.441.2923C} investigated the relationship between filament length and properties such as mass, diameter, and shape across a range of redshift, spanning from $z = 2$ to $z = 0$. More recently, \citet{2024arXiv240211678W} find that filament has a characteristic radius around 1Mpc/h. While many properties of filaments have been studied, little is currently known about the information filaments carry about the underlying cosmological model or how initial conditions affect the formation and evolution of filaments.

Over the past few years, numerous studies have explored various statistical methods to extract cosmological information, surpassing simple power spectrum analyses. These methods include incorporating information from higher-order statistics \citep{2020JCAP...03..040H,2021JCAP...04..029H,2021JCAP...07..008G,2022PhRvD.106d3530P}, velocities \citep{2015ApJ...808...47M,2021A&A...653A.130K,2022A&A...660A.113K}, marked two point correlation function and power spectrum \citep{2021PhRvL.126a1301M,2022arXiv220601709M,2021JCAP...03..038P,2020ApJ...900....6Y,2022MNRAS.513..595X,2019MNRAS.485.5276F}, tomographic Alcock-Paczynski (AP) \citep{2014ApJ...796..137L,2015MNRAS.450..807L,2016ApJ...832..103L,2018ApJ...856...88L,2019ApJ...887..125L}, wavelet scattering transform \citep{2020PhRvD.102j3506A,2020MNRAS.499.5902C,2021arXiv211201288C,2022PhRvD.106j3509V,2022PhRvD.105j3534V,2022arXiv220407646E},continuous wavelet transform \citep{2022ApJ...934..112W,2022ApJ...934...77W}, split densities \citep{2020MNRAS.495.4006U,2021MNRAS.505.5731P}, partial or total cosmological environments with two point correlation function or power spectrum
\citep{2005MNRAS.357..608Y,2012MNRAS.421.3481L,2015MNRAS.451.1036C,2015JCAP...11..036H,2015PhRvD..92h3531P,2016PhRvL.117i1302H,2017JCAP...07..014H,2019JCAP...12..040V,2022ApJ...935..100K,2021ApJ...919...24B,2022MNRAS.516.4307W,2022arXiv221206838B,2022A&A...661A.146B,2023MNRAS.tmp..989P}, cluster density function \citep{2009ApJ...692.1060V,2010MNRAS.406.1759M,2012ApJ...708..645R,2016A&A...594A..24P,2019ApJ...878...55B,2019MNRAS.488.4779C,2021PhRvD.103d3522C}.
These implications suggest that we can identify additional tracers to extract information about cosmology and the large scale structure of the universe. Filaments provide a promising avenue for studying the universe, and their properties can be used to constrain cosmological parameters and examine the initial conditions of the universe.
In this paper, we focus specifically on one property of filaments - their length distribution - to investigate how cosmology and initial conditions affect the formation and evolution of filaments. We explore the potential of filament length as a new tool for cosmological statistics in various ways. As ongoing and future surveys such as Dark Energy Spectroscopic Instrument \citep{2016arXiv161100036D}, Euclid \citep{2011arXiv1110.3193L}, the Nancy Grace Roman space telescope \citep{2015arXiv150303757S} and the Chinese Space Station Telescope (CSST) will be more powerful to construct cosmic filament from observations, our methods may have a valuable and promising application to constrain the cosmology parameters. Additionally, we present a simple test aimed at understanding how initial fluctuations influence filament formation, which can help uncover the hidden cosmological information embedded in the cosmic web.

The layout of this paper is as follows. In Sect.~\ref{sect:method}, we introduce the simulation data used in this study as well as the filament finder methods. Detail analysis of the filament length function is presented in Sect.~\ref{sect:result}, and finally the conclusion and discussion are present in Sect.~\ref{sect:conclusion}.

\section{Method}
\label{sect:method}

\subsection{Simulation dataset}
\label{sec:dataset} 

We utilize a large amount of simulations that was simulated using the COmoving Lagrangian Acceleration (COLA) code \citep{2013JCAP...06..036T}. COLA calculates the movement of dark matter particles in a comoving frame, following the trajectories forecasted by Lagrangian Perturbation Theory (LPT), enabling it to accurately handle small-scale structures while preserving accuracy on large scales. Despite being hundreds of times faster than N-body simulations, it maintains high precision from very large to highly non-linear scales. To explore various aspects of filament length statistics and evaluate the impact of numerical noise on analysis results and test robustness, we utilize COLA to generate 700 pure dark matter cosmological simulations with different parameters. 
The parameters of these simulation are listed in Table.\ref{tab:simulation_table} and they are categorized as follows,



\begin{enumerate}
\item[1)] 
In the fiducial dataset, we conduct 50 standard cosmological simulations with the cosmological parameters of Planck15, with $\Omega_m = 0.308$, $\Omega_b = 0.05$, $\sigma_8 = 0.83$, $n_s = 0.968$, and $H_0 = 67.8 , \text{km s}^{-1} \text{Mpc}^{-1}$. Each simulation box has a size of $512^3 (h^{-1}\text{Mpc})^3$ and contains $256^3$ particles.

\item[2)]
P1 and P2 datasets have the same cosmological parameters as the fiducial one, but with different number of particles: $128^3$ and $512^3$ respectively. These two datasets are used to investigate the impact of varying particle numbers on filament searches and statistics. Each dataset contains five realizations with different random seeds for producing initial conditions.

\item[3)]
The C1 dataset comprises 80 cosmological simulations, with varying $\Lambda$CDM parameters: $\Omega_m$, $\sigma_8$, and $n_s$. Our goal is to analyze changes in filament length statistics across various cosmological parameters and determine their potential for constraining these parameters. To achieve this, we conduct simulations where the value of a single parameter is varied with respect to the fiducial simulations. Each simulation utilizes the same box size and particle number as the fiducial dataset. To ensure robustness, we also generate five realization for each simulation using distinct random seeds. The cosmological parameters for the C1 dataset ranged from [0.2075, 0.4075] for $\Omega_m$, [0.60, 1.00] for $\sigma_8$, and [0.60, 1.20] for $n_s$.


\item[4)]
The SC1 dataset closely resembles C1, except for a box size of $300^3 (h^{-1} {\rm Mpc})^3$ and $512^3$ particles, encompassing 80 cosmological simulations. This dataset serves to assess the robustness of our findings across varying simulation box sizes and precision levels.

\item[5)]
The I1 dataset comprises 240 cosmological simulations which have modified initial power spectrum. The goal is to study how filament length statistics are affected by initial power spectrum. We vary the initial fluctuations at eight different scales, with each scale containing six different fluctuation amplitudes, with a total of 48 simulations. For each modified power spectrum, we generate five realizations with different random seeds, using the same box size and particle number as the fiducial simulation.

\item[6)]
The SI1 dataset closely resembles I1, except for a box size of $300^3 (h^{-1} {\rm Mpc})^3$ and $512^3$ particles, consisting of 240 cosmological simulations. This dataset serves to assess the robustness of our findings across varying simulation box sizes and precision levels.

\end{enumerate}

\begin{table*}
	\centering
    \caption{The adopted model parameters for the total 700 simulations produced by COLA.  The data is categorized into seven groups, to cover different cosmology, box size, particle number and power spectrum.}
	\label{tab:simulation_table}
	\begin{tabular}{lccccccr} 
		\hline
		 & fiducial & P1 & P2 & C1 & SC1 & I1 & SI1 \\
		\hline
		Box size [$(h^{-1}{\rm Mpc})^3$] & 512 & 512 & 512 & 512 & 300 & 512 & 300 \\
	  Particle number  & $256^3$ & $128^3$ & $512^3$ & $256^3$ & $512^3$ & $256^3$ & $512^3$ \\
		Cosmological parameters & Planck15 & Planck15 & Planck15 & Different & Different & Planck15 & Planck15\\
        power spectrum & CLASS & CLASS & CLASS & CLASS & CLASS & modify & modify \\
        Random seeds & 50 & 5 & 5 & 5 & 5 & 5 & 5 \\
        Total & 50 & 5 & 5 & 80 & 80 & 240 & 240  \\
         \hline
	\end{tabular}

\end{table*}

By running COLA, we gain access to a diverse range of cosmological simulations, enabling us to extract valuable physical insights and comprehend the intricate, nonlinear nature of cosmic filaments. This process aids in the development of more robust statistical methods for observational constraints on cosmological models.

\subsection{Filament pipline}
\label{sec:pipline}
After acquiring a cosmological simulation, we utilize the CIC (Cloud-in-Cell) scheme \citep{1998ARA&A..36..599B} to allocate dark matter particles onto a mesh grid, thereby generating density fields for specific snapshots similar to \citet{2022A&A...661A.115G}. In this work we only focus on the properties of filaments at five specific redshift, from z=0, 0.5,1,2,3. We then apply the filament finder, DisPerSE, to the density field to identify filaments and study their properties and dependence on cosmological parameters.

It is worth noting that, since we can only observe galaxies, filament finders should ideally be applied to the distribution of galaxies, as done by \citep{2020A&A...641A.173G,2021A&A...649A.117G,2022A&A...661A.115G,2023A&A...671A.160G,2024A&A...684A..63G}. One major concern in these studies arises from the need to construct the underlying dark matter density using galaxy distribution, which is inevitably affected by the uncertainty of galaxy bias. In our current studies, we aim to investigate the cosmological dependence of filament distribution by using the dark matter density field directly from simulations. This approach avoids any bias from the sparse tracers of the cosmic density field, such as galaxies or haloes. However, to extract constraints on cosmology, one will ultimately have to use observed galaxies to identify cosmic filaments, which will be a focus of future studies.

We employ the filament finder DisPerSE to extract filamentary structure from the DM density field. DisPerSE \citep{2011MNRAS.414..350S} is a publicly available code that identifies the critical points of the density field using the Morse method, where $S=4,3,2,1$ corresponds to bifurcation, node, saddle, and void, respectively. The saddle connects two nodes to form a filament \citep{2019MNRAS.483.3227K}, and bifurcation is where multiple filaments intersect, which is considered unresolved clusters and the location where superclusters exist. Building upon these considerations and inspired by the cluster-filament-cluster configuration proposed by  \citet{1996Natur.380..603B}, we define a complete filament as a high-density line connecting two nodes or bifurcations, or linking a cluster and a bifurcation which is the same as the definition proposed by \citet{2024arXiv240211678W}.

\begin{figure}
	\includegraphics[width=\columnwidth]{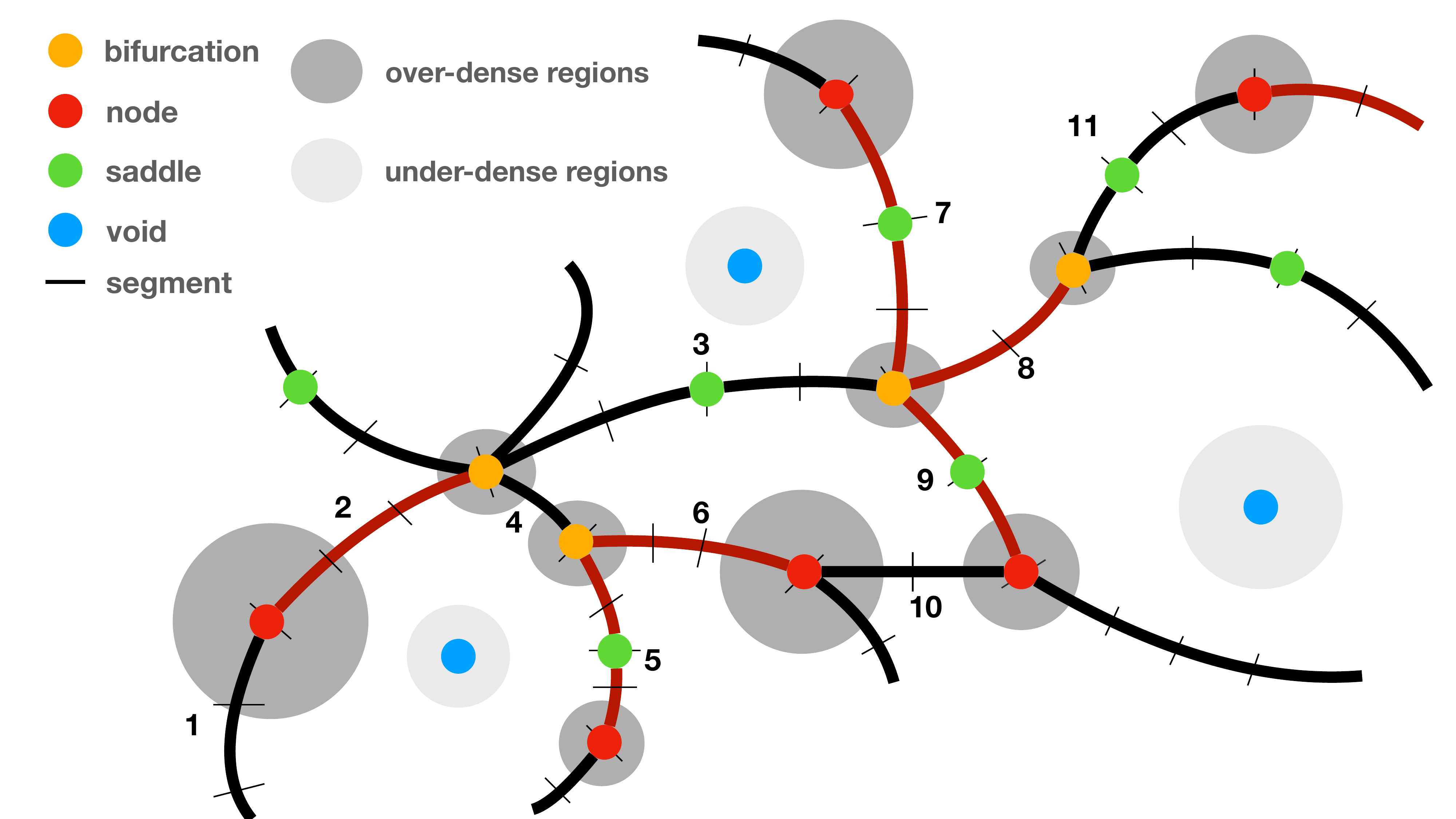}\\
   \caption{Similar to the figure by \citet{2020A&A...641A.173G}, the critical points (bifurcation, node, saddle, and void) and filament segments are identified using DisPerSE. Saddle points connect segments to form complete filaments. In our study, each complete filament connects two nodes or bifurcations, or links a cluster and bifurcation according to the cluster-filament-cluster configuration. We use black and brown colors to distinguish different filaments and numbers to tag each filament in the figure. There are a total of 11 complete filaments.}
    \label{fig:filament}
\end{figure}

DisPerSE segments the identified filaments into multiple parts, which are subsequently reconnected to form complete filaments based on the definition above. Fig.~\ref{fig:filament} illustrates the critical points and segments of filaments obtained using DisPerSE, similar to the figure by \citet{2020A&A...641A.173G}, although with a distinct definition of an individual filament. We use black and brown colors to distinguish different filaments and numbers to tag each filament in the figure, where there are a total of 11 complete filaments.

Ultimately, we define the length of a filament, denoted as 
$L$, as the sum of the lengths of the individual segments comprising the complete filament. This summation encompasses segments between nodes to nodes, nodes to bifurcations, or bifurcations to bifurcations. By computing the lengths of all complete filaments, we derive the filament length function.
Here, we define the number density of filaments in each length bin as $\Phi_{L}$.
\begin{equation}
    \Phi_L = \frac{N(L)}{V_{boxsize}}
\end{equation}
Here, $N(L)$ represents the number of filaments in each length bin, and $V_{boxsize}$ is the volume of the simulation box. Note that in this paper all quantitative relationships are derived with this number density $\Phi_L$, but one should be aware that cosmological fluctuation is significant when $N(L)$ is smaller. In addition, we define the relative difference of $\Phi_L$ as,
\begin{equation}
T=\frac{N(L)}{N(L)^{\prime}}-1
\end{equation}
to show the comparison of filament length distribution between different parameters. For example, $T(L)$ denotes the deviation of the filament length distribution respect to the reference value.

The filament identification by DisPerSE is influenced by several factors: the number of grid points, denoted as $N_{grid}$, the smoothing scale of the field, denoted as $N_{smooth}$, and topological persistence, denoted as $N_{\sigma}$. Given our emphasis on the cosmological dependence of filament length distribution, it is vital to ensure that our field structure is accurate and robust, enabling us to draw essential physical conclusions while minimizing the introduction of numerical noise.

Coarse fields and smaller $N_{grid}$ leads to the elimination of small-scale information and minor structural details. we have employed a boxsize of 512 $h^{-1}\text{Mpc}$ with 512-grid point setup, ensuring the accuracy on small scales and providing ample resolution for the cosmological scale under investigation.

On the other hand, the smoothing scale $N_{smooth}$ involves eliminating the shot noise introduced by the lattice process from point cloud to field. Following previous studies on the optimization of smoothing length \citep{2007MNRAS.375..489H,2014MNRAS.441.2923C}, we use a Gaussian smoothing with $N_{smooth} =2 h^{-1} {\rm Mpc}$ to get the density field. This smoothing effectively eliminates shot noise in the density field below a mass scale of $10^{13} M_\odot h^{-1}$ and filaments with lengths below $2 h^{-1} {\rm Mpc}$.

$N_{\sigma}$ characterizes topological persistence, aiding in the identification of robust components of the filamentary network in comparison to a discrete random Poisson distribution. 
To determine the optimal value for the topological persistence parameter $N_{\sigma}$, we employ the $CP_{max}-Halo$ calibration method to refine our pipeline settings. This method has been widely utilized in previous studies (e.g., \citep{2024A&A...684A..63G,2023A&A...671A.160G,2024arXiv240204837B}).
The primary concept of this method involves comparing the positions of nodes ($CP_{\text{max}}$) to those of the most massive halos identified by the Friend-of-Friends (FoF) algorithm \citep{1985ApJ...292..371D}, which is widely used to find dark matter halo by linking all particles into one halo when the distance between any two particles is less than f times the average distance between the particles in the simulation box, where f is usually taken as 0.2. The results are depicted in Fig.~\ref{fig:cph}. Following the approach of \citet{2024arXiv240204837B}, we present the ratio of the fraction of $CP_{\text{max}}$ located within the virial radius ($r_{\text{vir}}$) of any massive halo to the fraction of massive halos hosting a $CP_{\text{max}}$ within their $r_{\text{vir}}$ spheres. This comparison aims to identify the topological persistence value that brings the ratio closest to unity, thus minimizing both the number of halos lacking nodes association and the number of nodes lacking halo association at $z = 0$. Consequently, we select $N_{\sigma} = 3$ which corresponds to the ratio we defined as unity. Notably, we designate the most massive halos as those with masses exceeding $10^{14} M_{\odot}h^{-1}$ to ensure compatibility with the smoothing scale encompassing $10^{13} M_{\odot}h^{-1}$. Detail analysis of the choosing $N_{\sigma}$ can be found in Appendix.\ref{sec:nsigma}.


This complete pipeline, starting from the initial power spectrum and given different cosmological parameters, allows us to obtain the statistics of filament lengths. A short summarize of the pipeline involves the following steps,

\begin{enumerate}
\item[1)]
Input the initial power spectrum and cosmological parameters, then employ COLA to generate snapshots at various redshifts.
\item[2)]
Utilize the Cloud-in-Cell (CIC) method on the snapshot acquired at a specific redshift to derive a smoothed density field.
\item[3)]
Employ the DisPerSE method to extract the diverse components of filaments.
\item[4)]
Construct a complete list of filaments according to the cluster-filament-cluster definition.
\item[5)]
Obtain various statistical measures for each filament.

Therefore, we can utilize this pipeline to examine numerous filament statistics, and the pipeline indicates that the statistics are solely dictated by the initial power spectrum and cosmological parameters. 


\end{enumerate}


\begin{figure*}
	\includegraphics[width=0.32\textwidth]{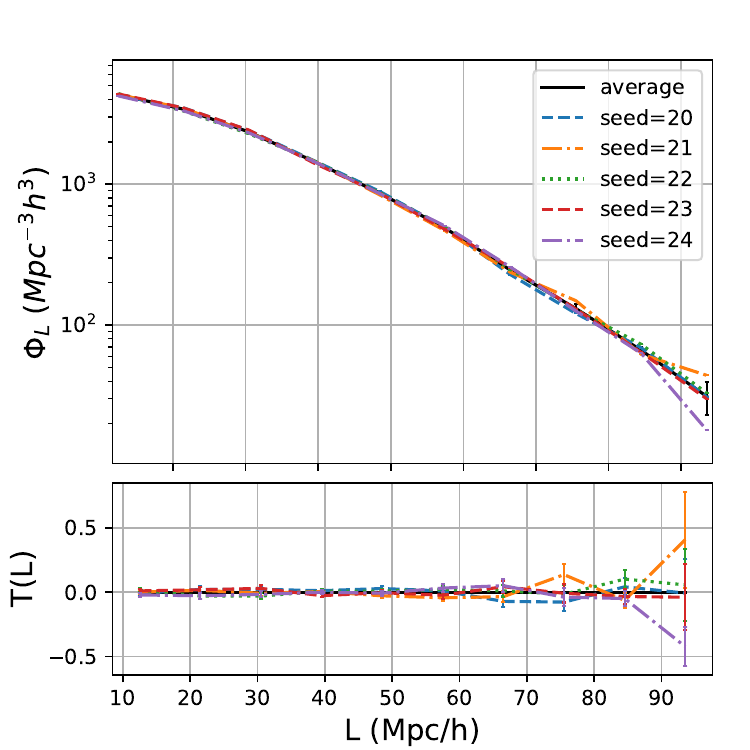}
    \includegraphics[width=0.32\textwidth]{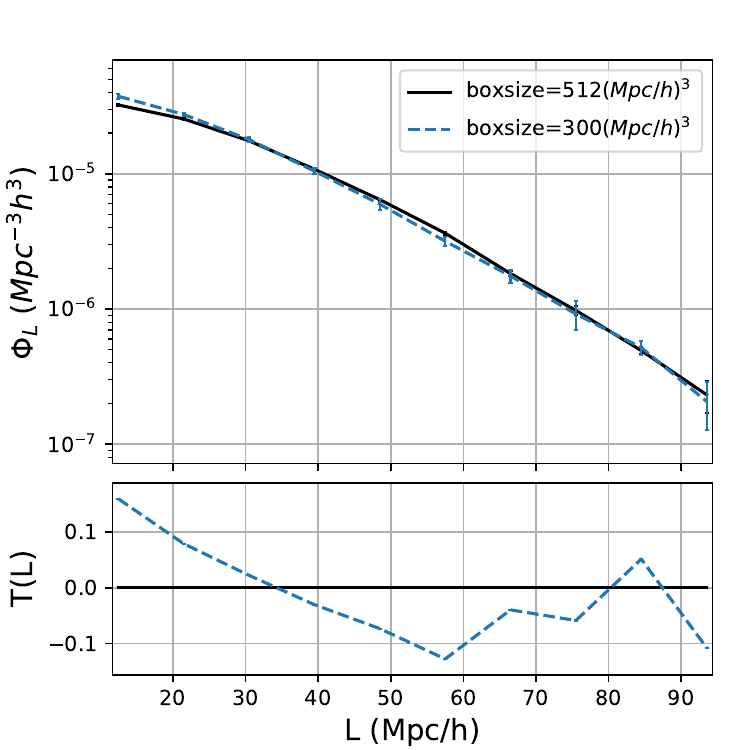}
    \includegraphics[width=0.32\textwidth]{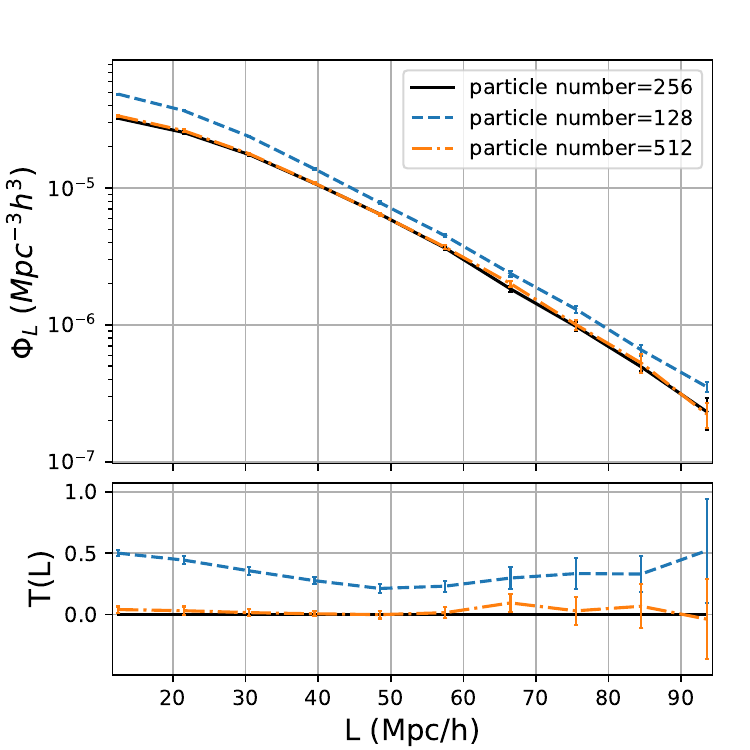}
   \caption{Robustness test for the filament length functions. From the left to the right panels, it shows the cosmic variance due to random seed for simulation realization, boxsize and particle numbers. See text for more details.}
    \label{fig:cosmic_variance}
\end{figure*}

\begin{figure*}
\centering
	\includegraphics[width=0.32\textwidth]{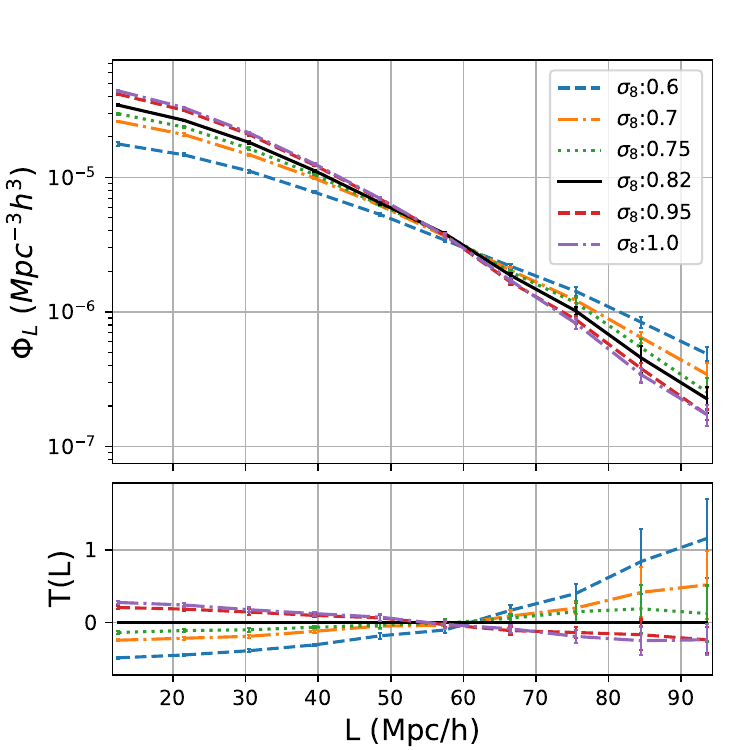}
        \includegraphics[width=0.32\textwidth]{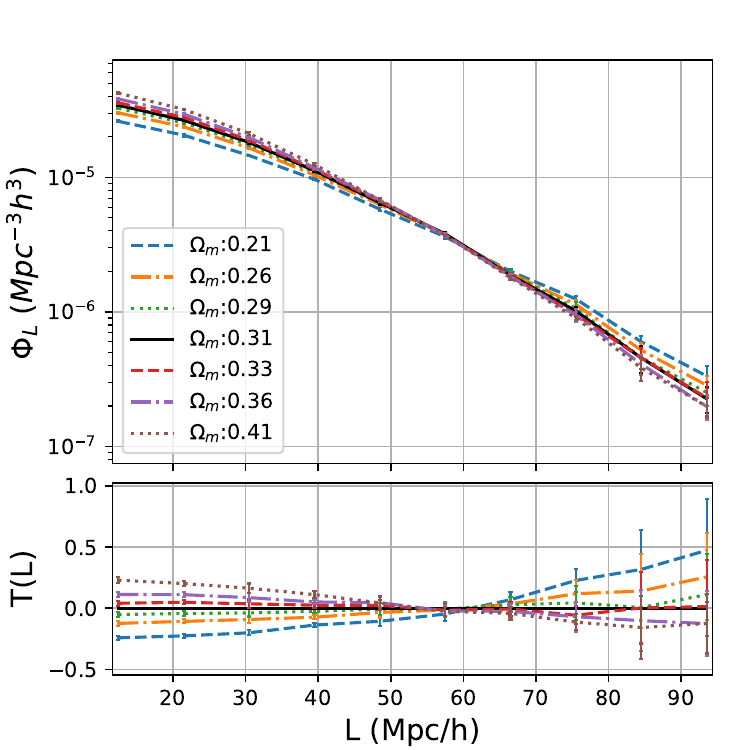}
        \includegraphics[width=0.32\textwidth]{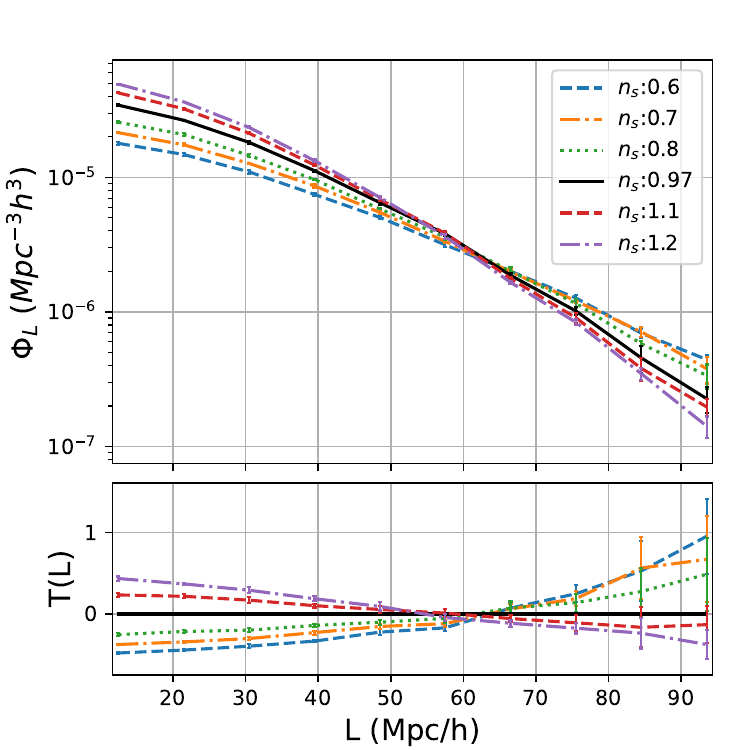}
   \caption{Filament length distributions for different cosmological parameters. The left, middle, and right panels present the differences of filament evolution functions for varying $\sigma_8$, $\Omega_m$, and $n_s$, respectively. For each cosmology, we have calculated the filament length distribution from five different initial matter distributions. Our findings reveal that the filament length function varies with changing cosmological parameters, validating filament length function as a valuable cosmological probe.}
    \label{fig:cosmology}
\end{figure*}

\begin{figure*}
\centering
	\includegraphics[width=0.32\textwidth]{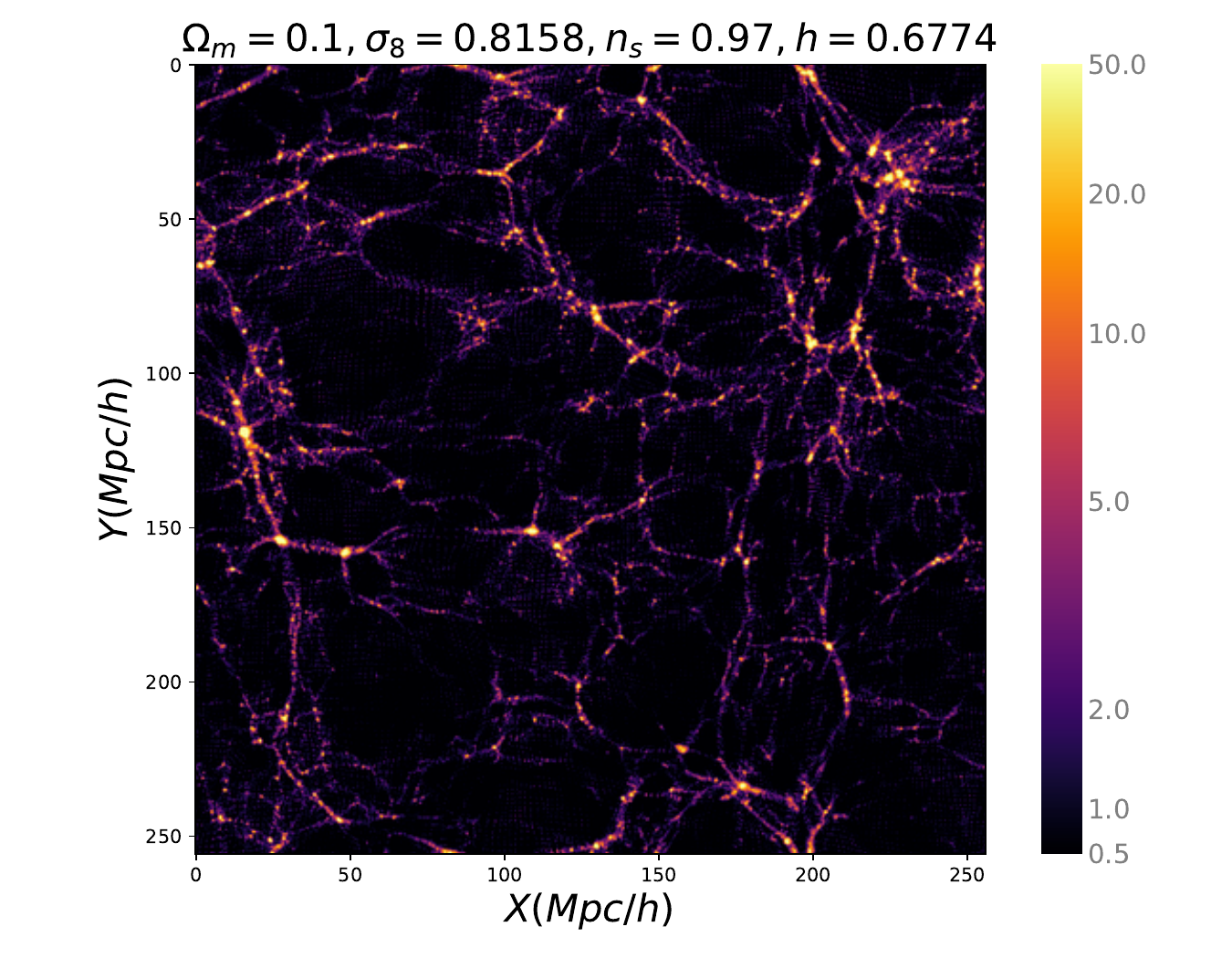}
    \includegraphics[width=0.32\textwidth]{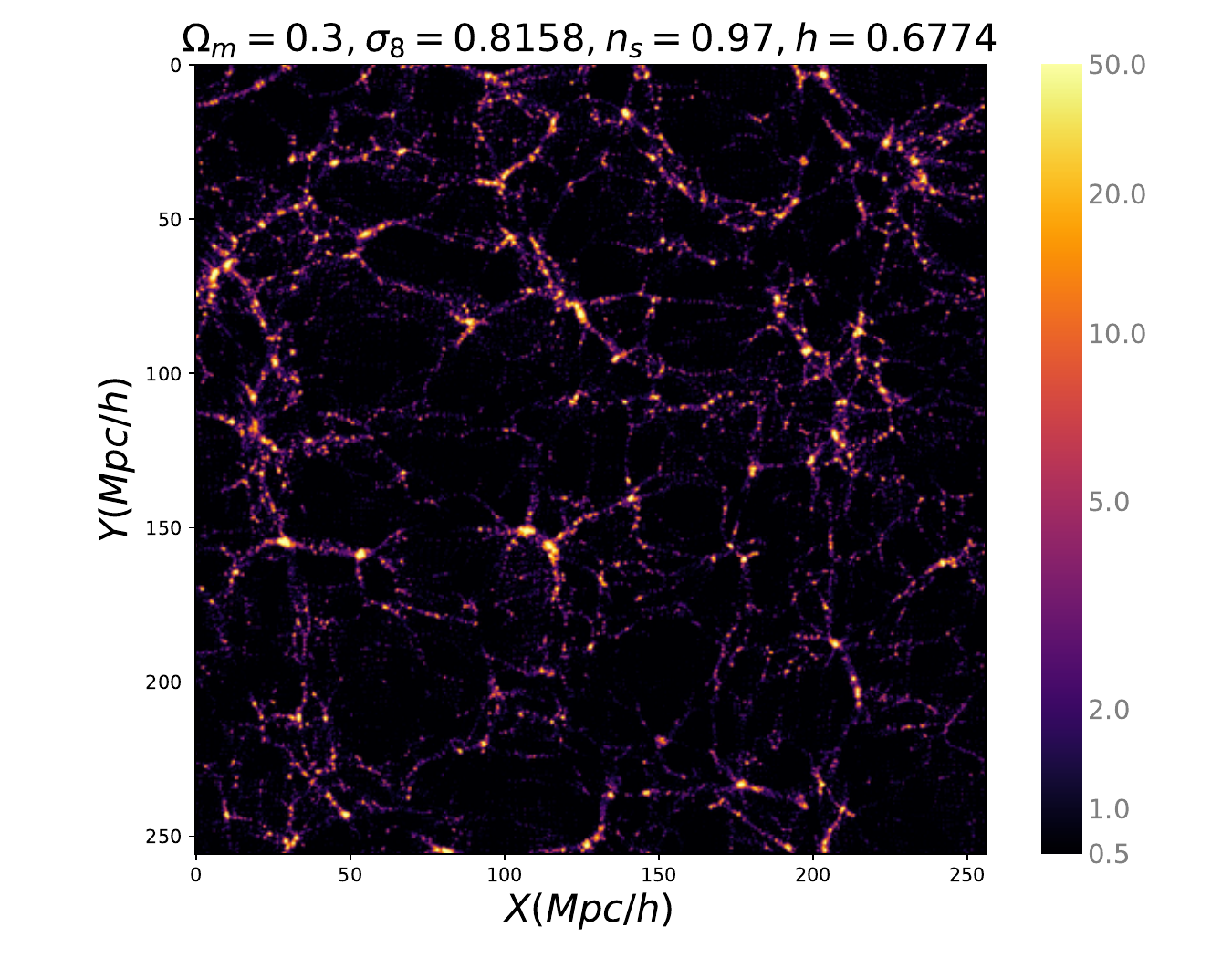}
    \includegraphics[width=0.32\textwidth]{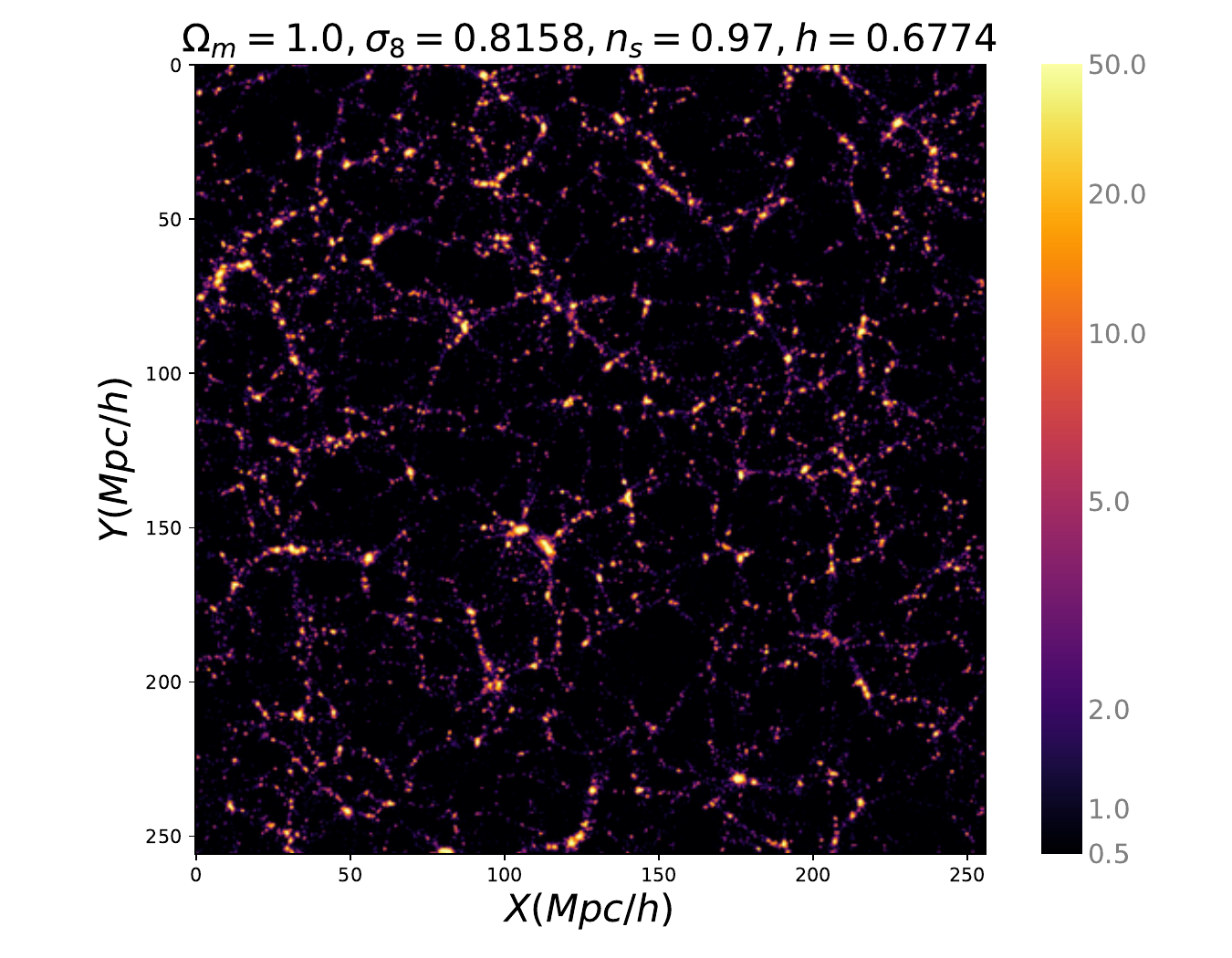}
   \caption{Slices of simulation output at $z=0$ for different $\sigma_8$. From the left to the middle panel, one can find that increasing $\sigma_8$ leads to development of more filament and more clusters, but the right panel shows that a too strong $\sigma_8$ leads to fragment of long filament into short ones. See text for more details.}
    \label{fig:pm_s8}
\end{figure*}

\begin{figure}
	\includegraphics[width=\columnwidth]{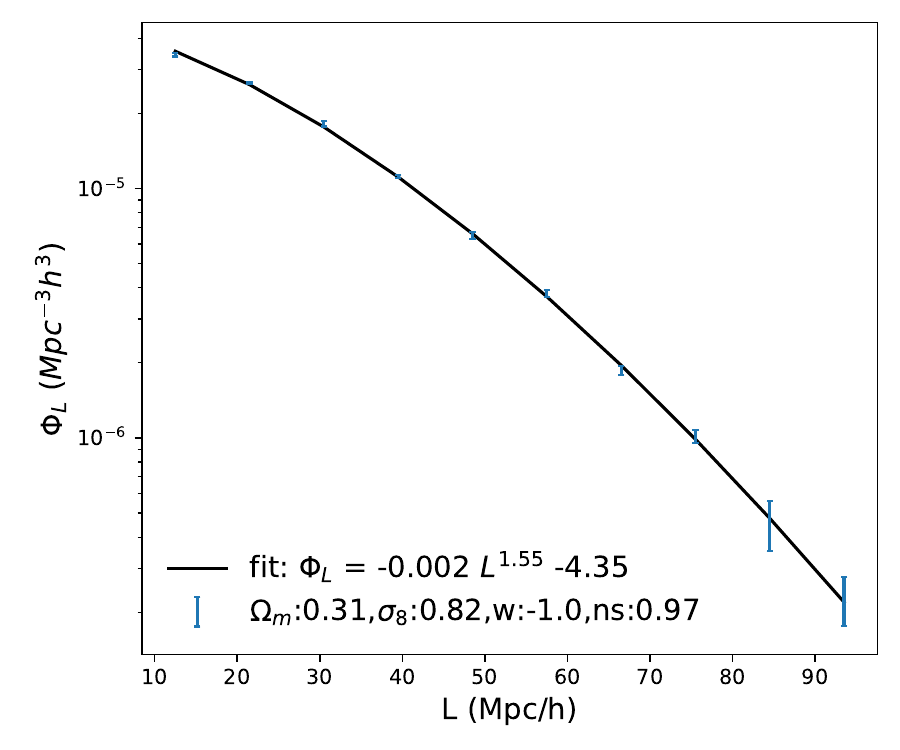}
   \caption{The filament length distribution for the Planck 15 cosmology at $z=0$ generated from five random seed values, represented by the blue dot with one sigma of variance. The black line shows that a power law fitting using Eq.\ref{eq:filament_fit} can well describe the filament length distribution.}
    \label{fig:fit_filament_length}
\end{figure}

\begin{figure*}
\centering
	\includegraphics[width=0.32\textwidth]{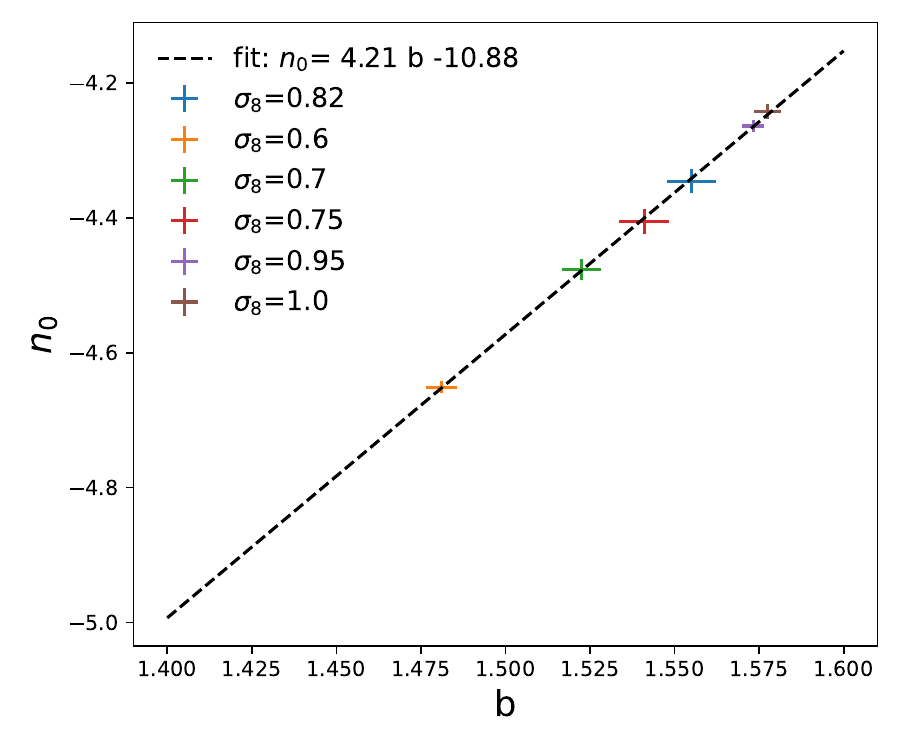}
        \includegraphics[width=0.32\textwidth]{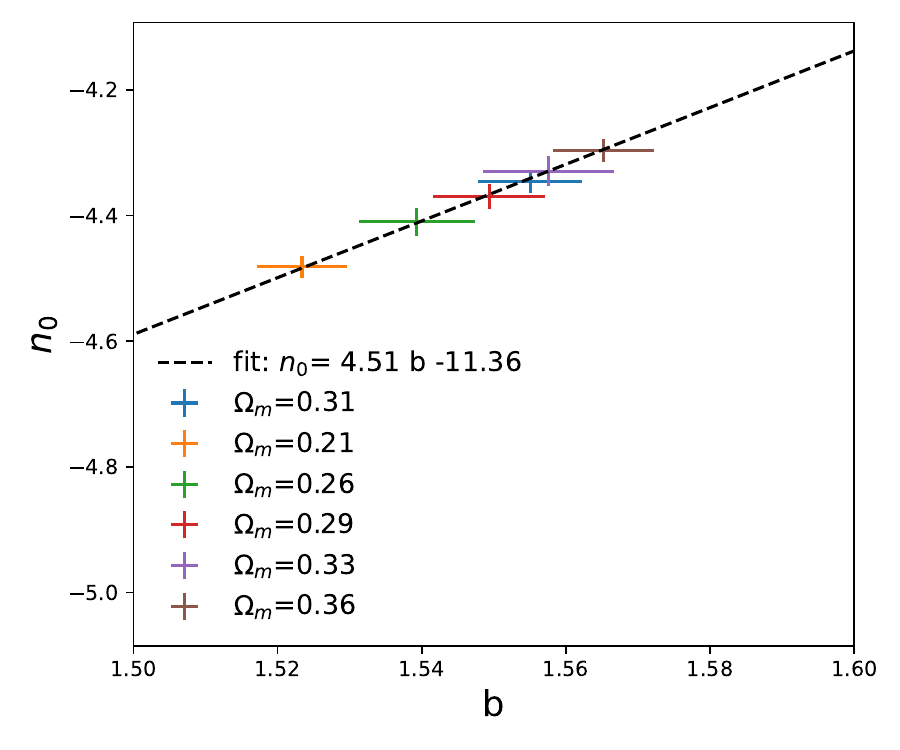}
        \includegraphics[width=0.32\textwidth]{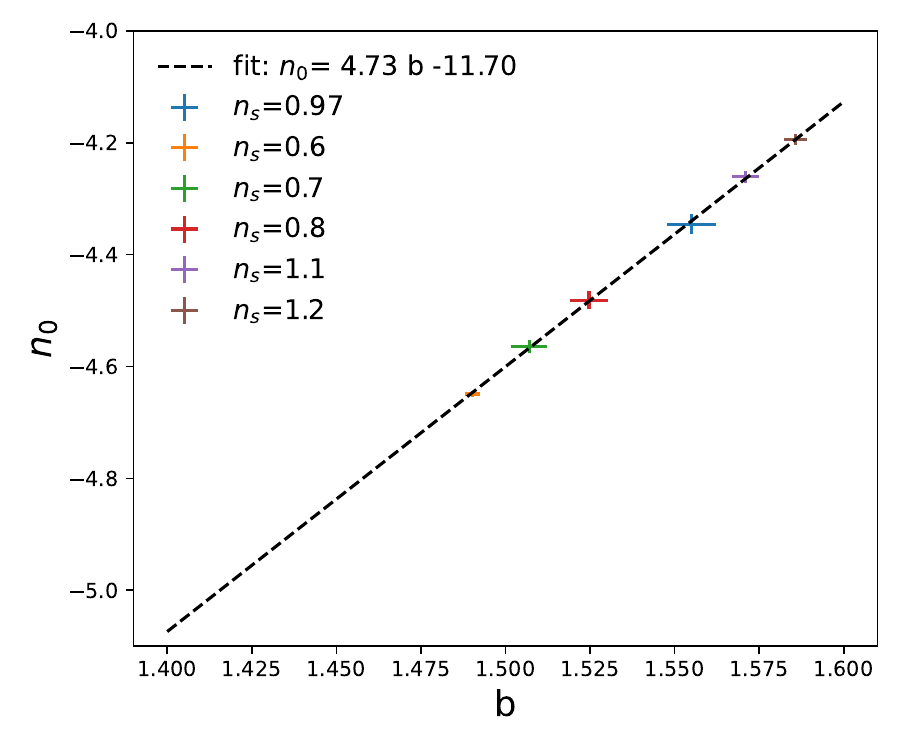}
   \caption{The dependence o $n_0$ and $b$, used to describe the filament length distribution, on the cosmological parameters. In each panel, five realization of the simulation is used to get the one sigma variance. It is seen that the filament length distribution has stronger dependence on $\sigma_8$. and $n_s$, with mild dependence on $\Omega_m$.}
    \label{fig:fit_filament_length_diffcosmo}
\end{figure*}

\begin{figure*}
\centering
	\includegraphics[width=0.32\textwidth]{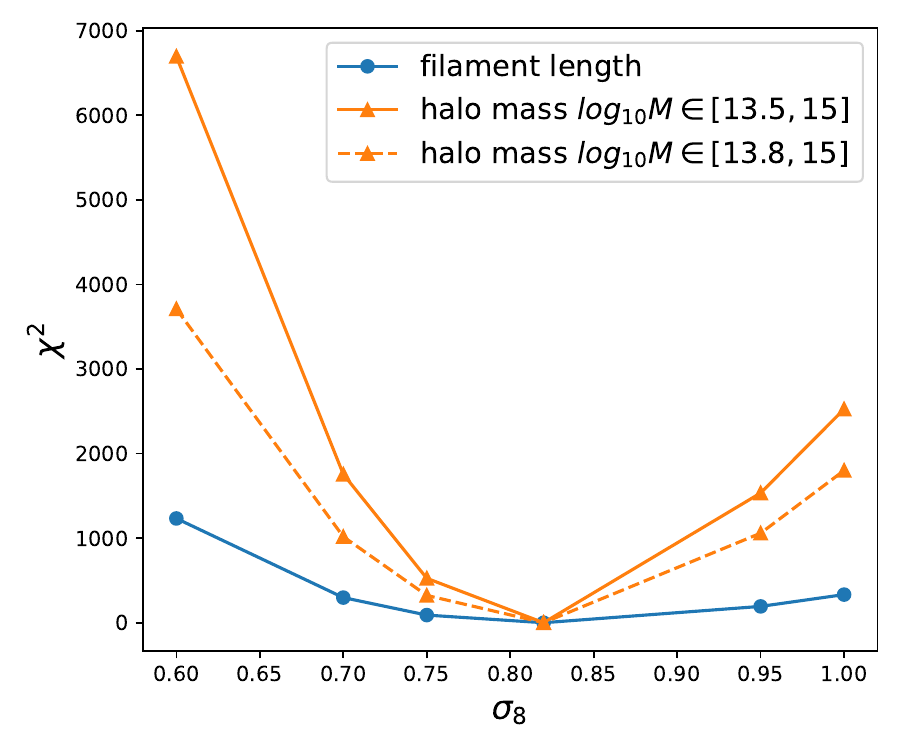}
        \includegraphics[width=0.32\textwidth]{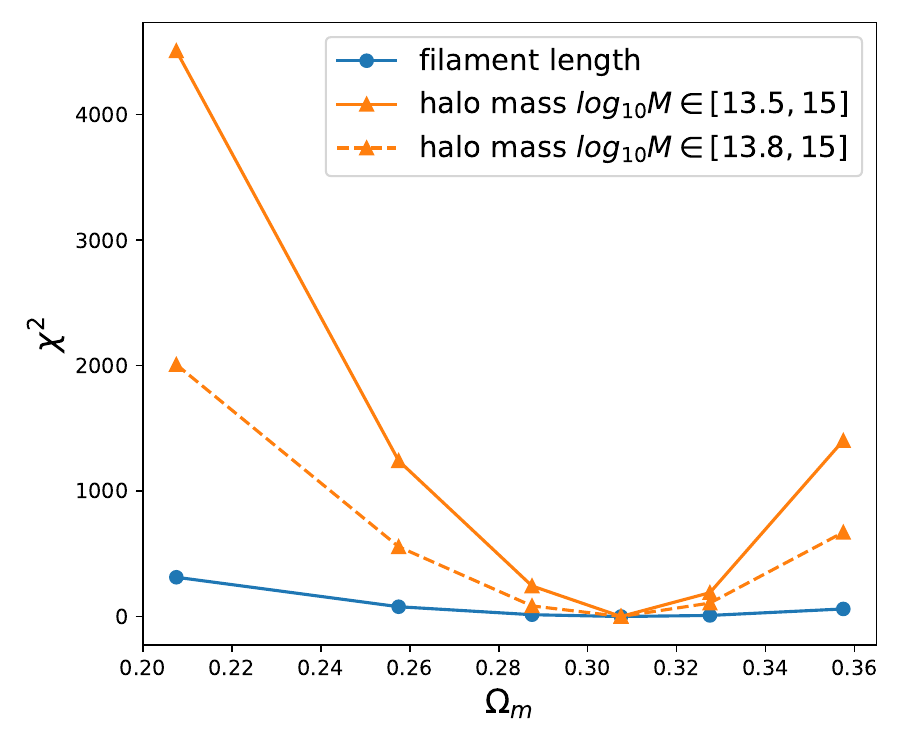}
        \includegraphics[width=0.32\textwidth]{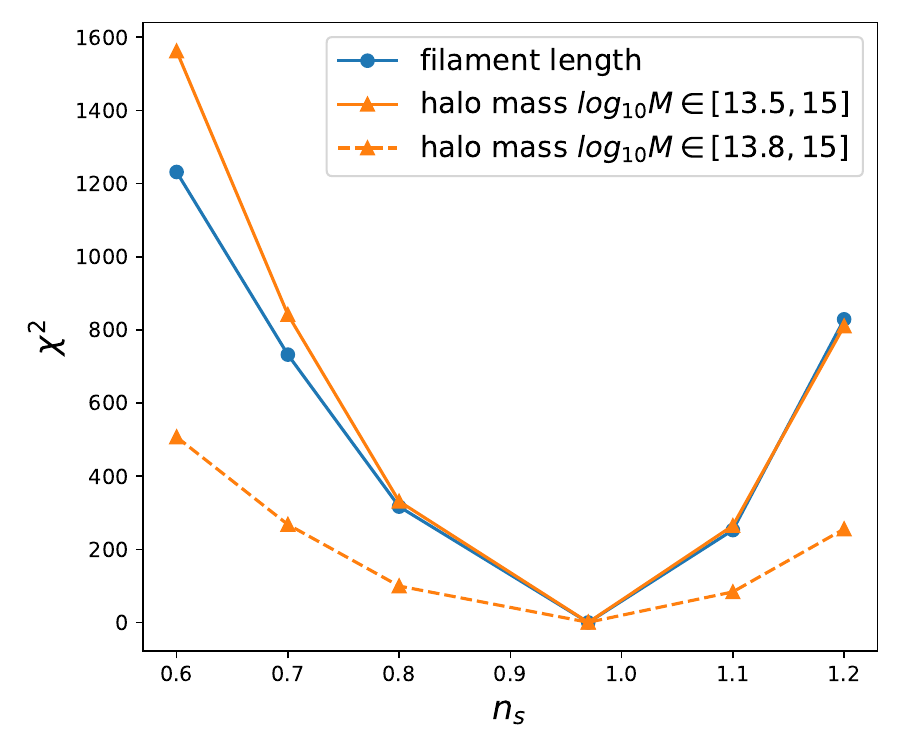}
   \caption{Conditional constraints on $\sigma_8$, $\Omega_m$ and $n_s$, derived using halo mass function and filament length function. Estimation is based on the simulation of 118 COLA simulation with a boxsize of 512 $(h^{-1} {\rm Mpc})^3$. The solid orange lines are for halo mass within $log_{10} M \in [13.5, 15]$, and dashed lines are for halo with slightly higher mass with $log_{10} M \in [13.8, 15]$. The result indicates that the filament length function is slightly less stringent in terms of cosmological constraints compared to the halo mass function.}
    \label{fig:x2_filament_length}
\end{figure*}

\begin{figure}

	\includegraphics[width=0.95\columnwidth]{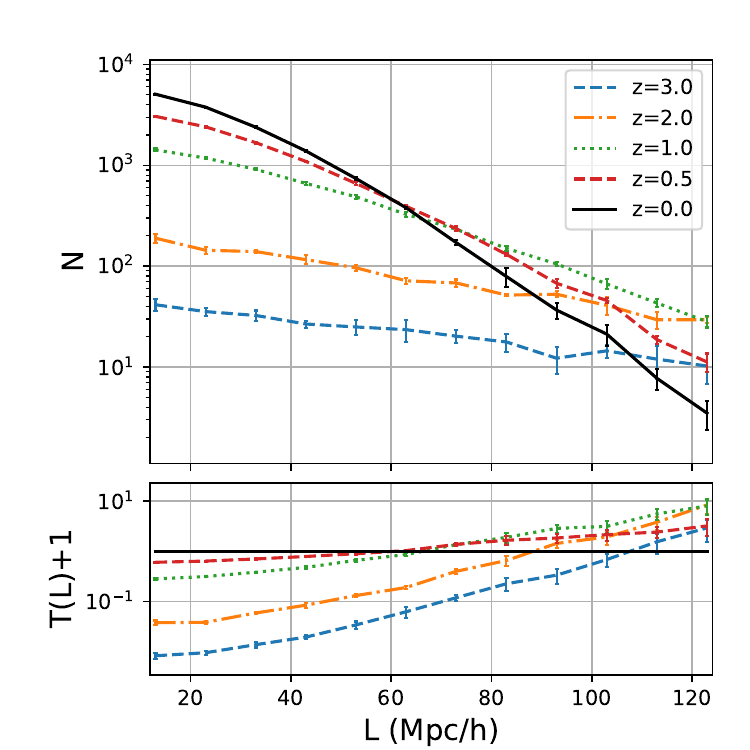}
   \caption{Filament length distributions at different redshift as well as their ratio to $z=0$ in Planck15 cosmology.}
    \label{fig:redshift}
\end{figure}

\begin{figure*}
\centering
	\includegraphics[width=0.45\textwidth]{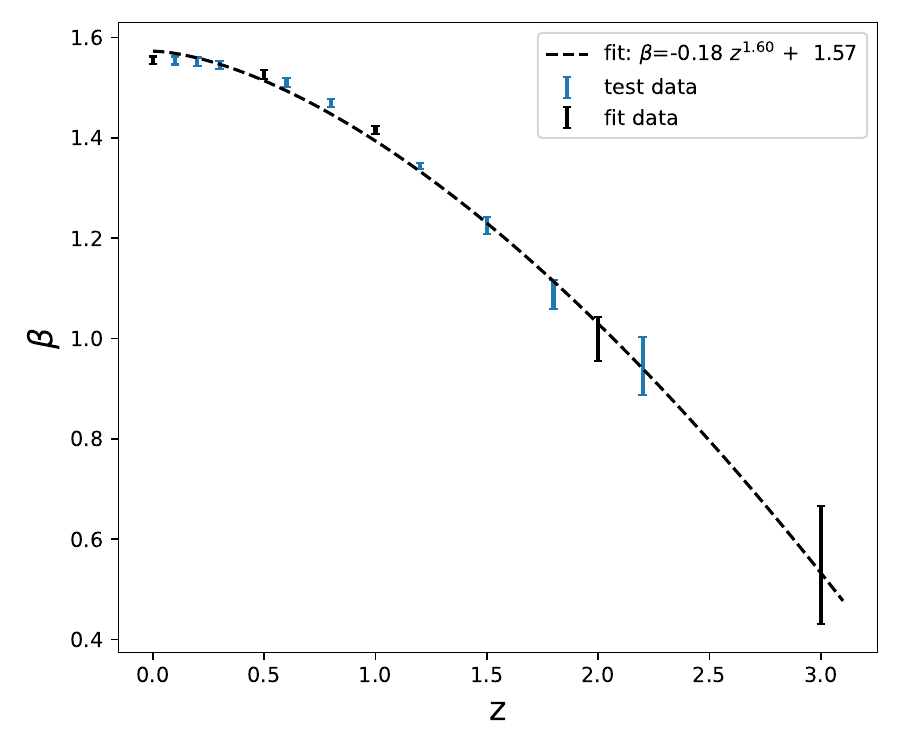}
 \includegraphics[width=0.45\textwidth]{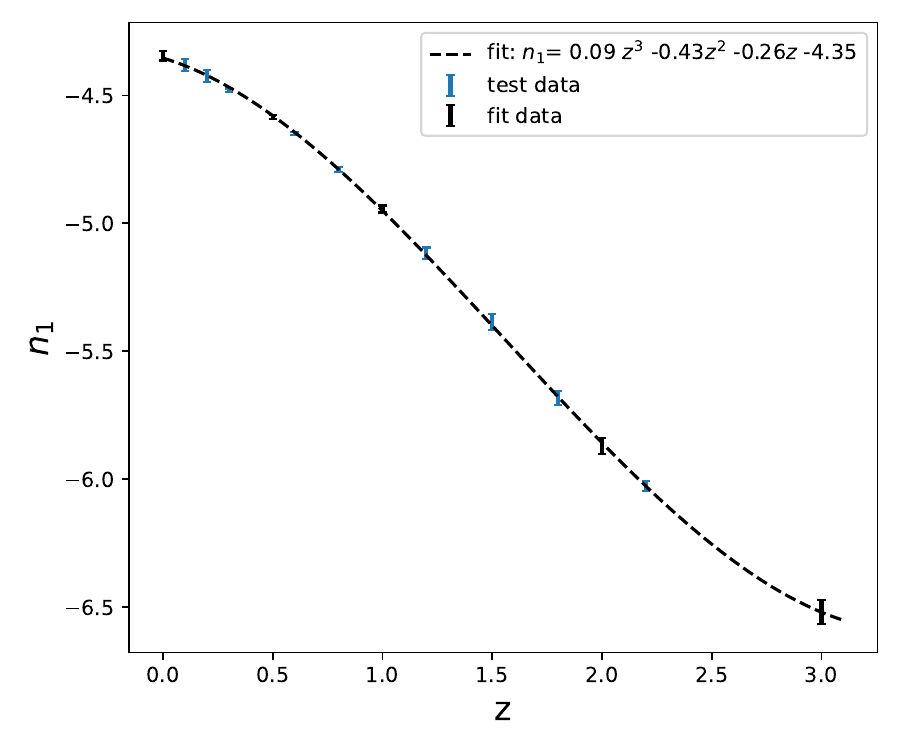}
   \caption{The relationship between the power law index $\beta$ and $n$ in the filament length function and redshift can be effectively described by the equation $\beta = a z^t + b$ and a cubic function, respectively. The black dots represent the 5 redshifts used for fitting this relationship, while the blue dots represent the redshifts that can be accurately predicted by the fitting relation.}
    \label{fig:beta_redshift}
\end{figure*}


\begin{figure}
\centering
	\includegraphics[width=0.50\textwidth]{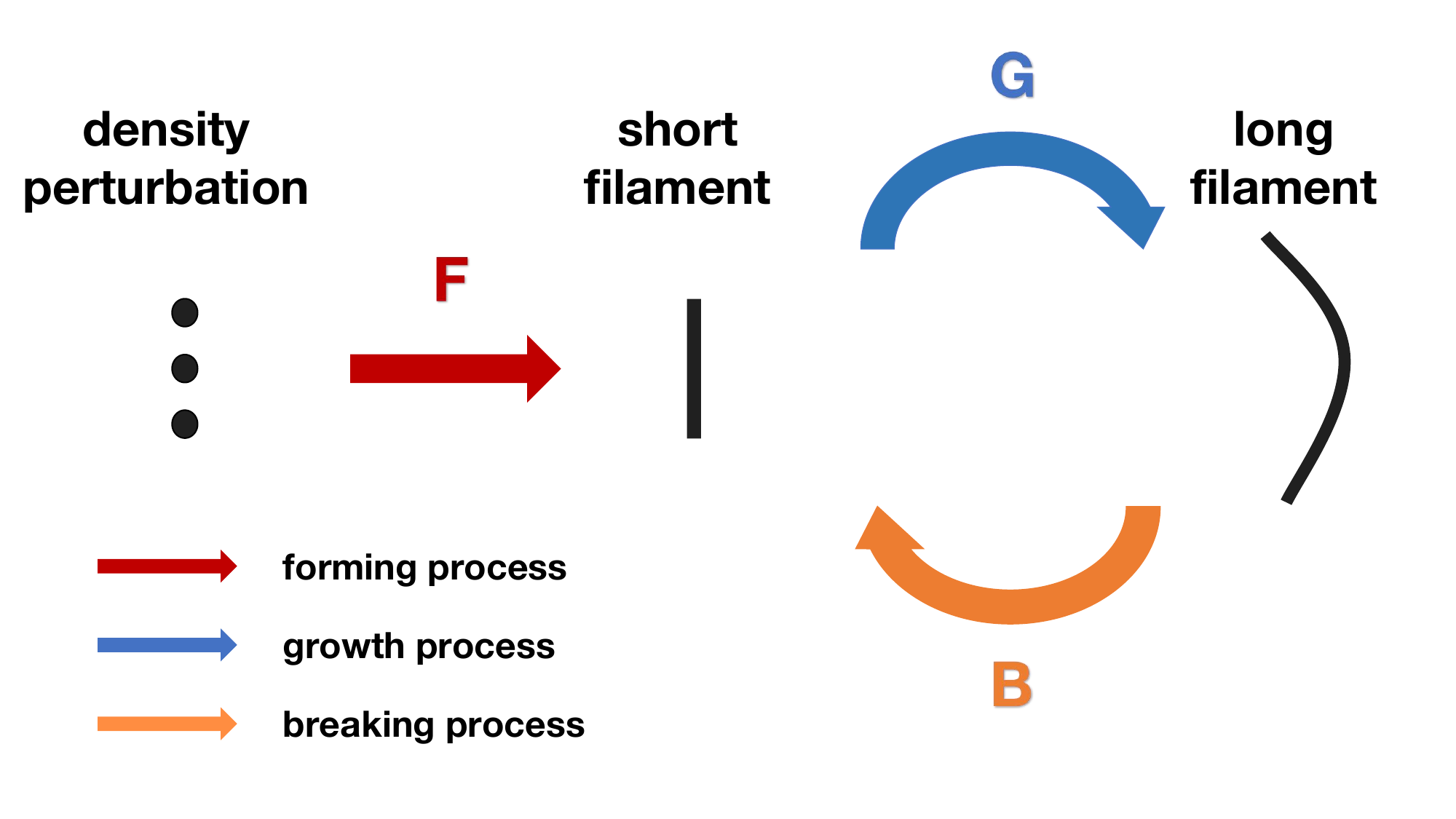}
   \caption{A cartoon plot to depict the formation and transformation of short and long filaments. The (F) process corresponds to the filaments forming process caused by density perturbations, and (B) represents the breaking down process from long filaments to short filaments. In addition, (G) represents the growth process from short filaments to long filaments.}
    \label{fig:filament_model}
\end{figure}

\begin{figure}
	\includegraphics[width=\columnwidth]{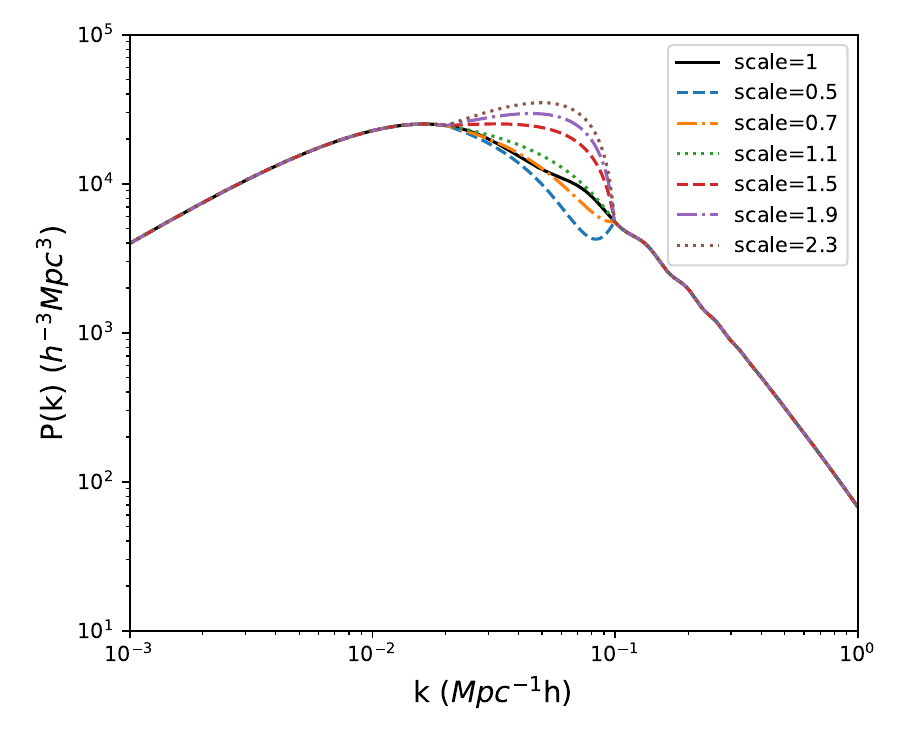}
   \caption{The initial power spectrum we generate in one of the intervals $k \in [0.02,0.1]$ with different amplitudes. }
    \label{fig:init_power_spectrum}
\end{figure}

\begin{figure*}
\centering
        \includegraphics[width=0.30\textwidth]{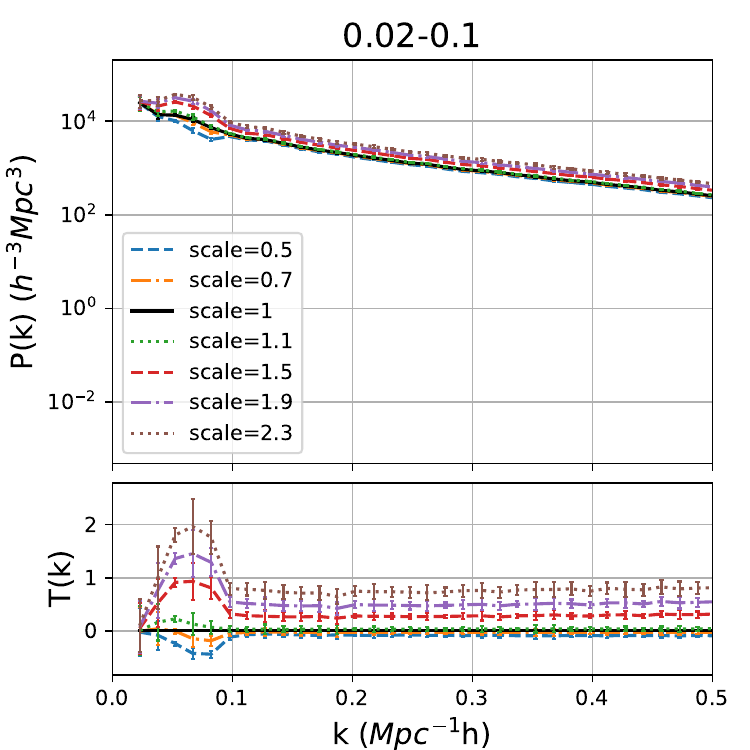}
        \includegraphics[width=0.30\textwidth]{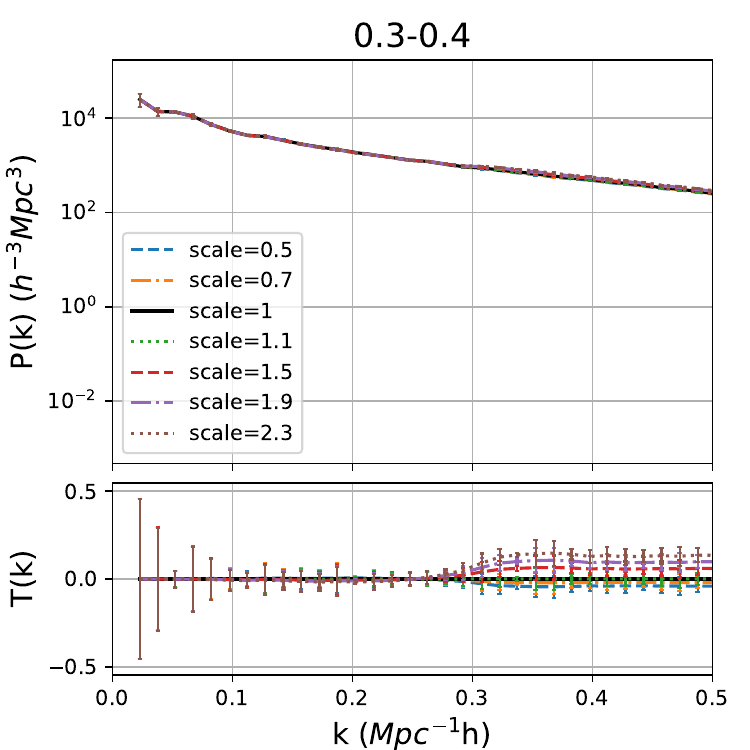}\\
        \includegraphics[width=0.30\textwidth]{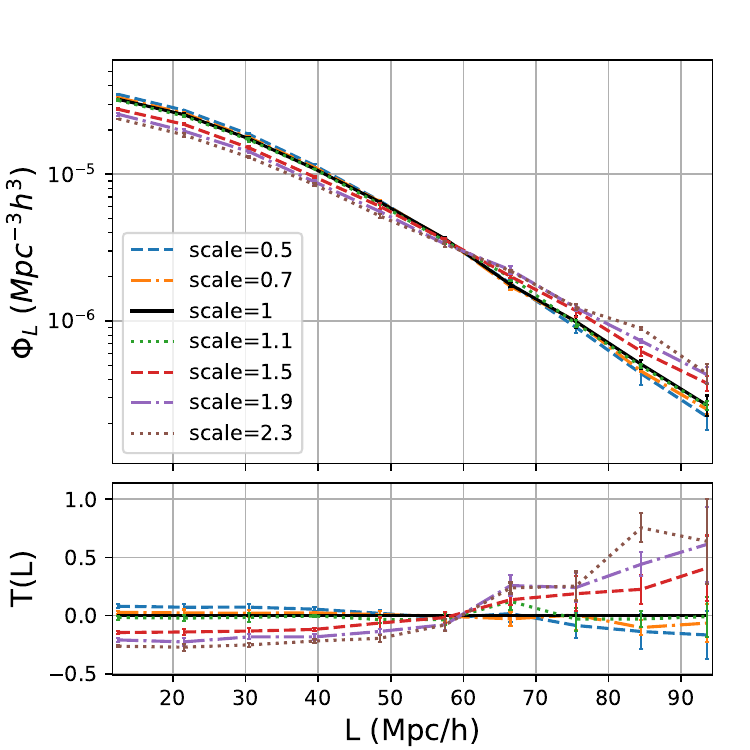}
        \includegraphics[width=0.30\textwidth]{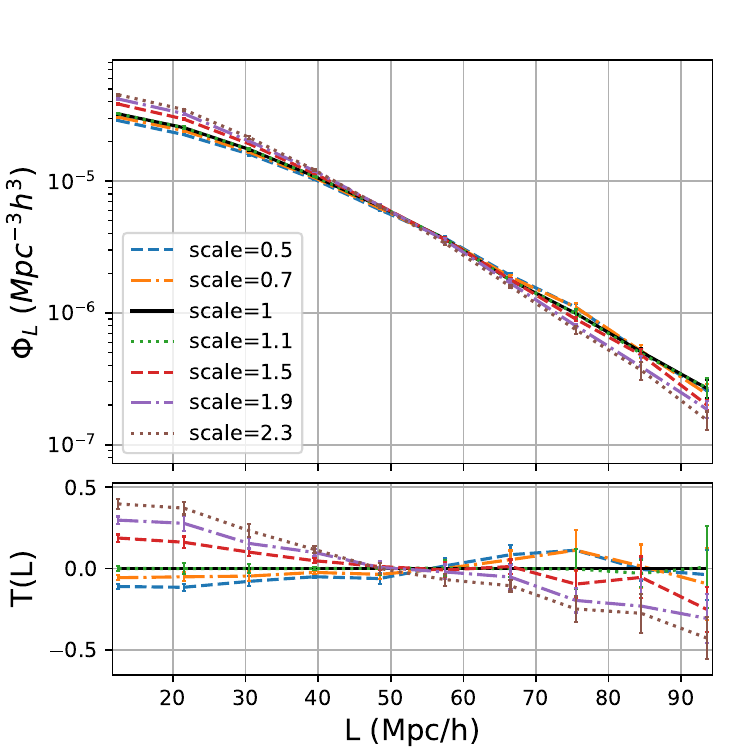}
        
   \caption{Comparison of the modified power spectra at $z=0$ and the filament length function. The fluctuations at different scales, specifically $k \in [0.02,0.1]$ and $k \in [0.3,0.4]$, result in changes in the formation of filaments and the large-scale structure. The first row displays the obtained power spectra, with each column corresponding to a different scale modification. The second row exhibits the corresponding filament length distribution, illustrating how modifications to the initial power spectrum affect the formation of filaments.}
    \label{fig:init_flutuation}
\end{figure*}

\section{Results}
\label{sect:result}

In this section, we will focus on the distribution and evolution of filament length, delving into its cosmological implications and its relevance to cosmology. 
In Section \ref{sec:cosmic_variance}, we test the robustness of the DisPerSE on our simulation setup, such as random seed, cosmic variance, grid number and smooth length. In Section \ref{sec:cosmic_parameter}, we present the length distribution of filament, with focuse on the dependence on cosmological parameters. In Section \ref{sec:redshift_evolution}, we study the redshift evolution of filament length distribution. In Section \ref{sec:initial}, we test how the shape of the initial power spectrum
affects the filament length distribution.

\subsection{Cosmic variance}
\label{sec:cosmic_variance}

In N-body simulation, one have to sample the power spectrum of the underlined cosmological density field, thus the final mass distribution is inevitable to be affected by the parameters used in N-body simulation, such as the random seed, particle number and box size of the simulation. Here we analyze their impacts one by one. 
In our fiducial simulation, we have 5 realization of a give cosmological parameters, namely with $\Omega_m = 0.308$, $\Omega_b = 0.05$, $\sigma_8 = 0.83$, $n_s = 0.968$, and $H_0 = 67.8{\rm km s}^{-1} {\rm Mpc}^{-1}$
In the left panel of Fig.~\ref{fig:cosmic_variance}, we show the filament length distributions of these realization. As can be seen from the left panel, the cosmological fluctuation of the filament length distribution is very small, with only a small scatter in filaments longer than $85h^{-1}{\rm Mpc}$.

To assess the influence of box size on filament length statistics, we examined the normalized filament length distribution using simulations with box sizes of $300 (h^{-1} {\rm Mpc})^3$ from the SC1 dataset and $512 (h^{-1} {\rm Mpc})^3$ within fiducial dataset, as depicted in the middle panel of Fig.~\ref{fig:cosmic_variance}, with 5 random seeds for each box size. Our analysis reveals that the variance resulting from box size is negligible and does not affect our results significantly. In principle, a larger box is preferable for analysis as it provides more filament samples and longer filaments. However, due to limited computing resources, we opt for a $512 (h^{-1} {\rm Mpc})^3$ box, which is sufficient for our current study.

In the right panel of Fig.~\ref{fig:cosmic_variance}, we compare the effects of varying particle numbers $N_p$ in numerical simulations on filament length statistics. Within a $512 (h^{-1} {\rm Mpc})^{3}$ box, the distribution of filament lengths exceeding $8 h^{-1} {\rm Mpc}$ is similar for particle numbers of $256^3$ and $512^3$, while simulations with $128^3$ particles differ significantly. This is because a low number density in the simulation results in a different distribution of the CIC density field. Therefore, to ensure both the accuracy of the density field and the efficiency of the simulation, we use $N_p = 256^3$.

The above results indicate that cosmological fluctuations including different random seed, different boxsize and different particle numbers do not significantly influence the filament length distribution. This outcome reinforces the viability of utilizing filament length distribution as a method for studying cosmological constraints.


\subsection{Cosmological parameters}
\label{sec:cosmic_parameter}

As mentioned in Section \ref{sec:Introduction}, the large-scale mass distribution is significantly affected by the cosmological parameters, as does the cosmic filament. However, few studies have focused on the effects of cosmology on filament length distribution. Using the filament pipeline discussed in Sec.\ref{sec:pipline}, we have investigated the relationship between filament length distribution and three different cosmological parameters in Fig.~\ref{fig:cosmology} where we show the results by varying $\sigma_8$, $\Omega_m$ and initial power spectrum index $n_s$, respectively. Overall, it is found that filament length distribution exhibits varying degrees of sensitivity to these three cosmological parameters, and the strength of this signal is significantly beyond the cosmic variance.  

In the left panel of Fig.~\ref{fig:cosmology}, we observe that as $\sigma_8$ increases, the number of filaments, particularly shorter ones, increases, while the number of longer filaments decreases. The decrease of $\sigma_8$ causes the universe to become more homogeneous, while an extremely low value of $\sigma_8$ would leave the filament structure nearly empty. As $\sigma_8$ increases and the clustering of each scale strengthens, more voids and peaks emerge, leading to an increase in the number of short filaments. However, as $\sigma_8$ reaches higher values, strong clustering results in increased merging, causing long filaments to break and additional peaks to form within them. This phenomenon may lead to the fragmentation of longer filaments into smaller ones, thereby contributing to an elevated number of short filaments.
To demonstrate this, we generate two-dimensional slice plots of the density field and employ the color rendering method from $pmwd$ \citep{2022arXiv221109958L} to effectively depict the cosmic structures of $\sigma_8 = 0.59$, $\sigma_8 =0.82$, and $\sigma_8 = 1.66$ at $z=0$ in a $256^3 (h^{-1} {\rm Mpc})^3$ box. This is illustrated in Fig.~\ref{fig:pm_s8}, where the absence of filamentary structure at low $\sigma_8$ is notable on the left panel. As $\sigma_8$ increases, more peaks form and the filament structure becomes apparent. However, with further increases in $\sigma_8$ to much higher values, more structures merge, leading to the break of long filaments and the emergence of additional peaks that break down the filaments. Consequently, the number of long filaments decreases while the number of short filaments increases.

Seen from the middle panel of Fig.~\ref{fig:cosmology}, it is evident that as $\Omega_m$ increases, short filaments increase while long filaments diminish. This phenomenon is explicable: the greater proportion of matter fosters nearby merging, each culminating in a small peak, thereby impeding the formation of extended structures.

While an increase in $n_s$ exhibits similar behavior to $\Omega_m$, it results in a stronger change in filament length distribution, as shown on the right panel of Fig.~\ref{fig:cosmology}.
The features are very similar to those of $\Omega_m$, with extensive structures present at small $n_s$, and as $n_s$ increases, smaller structures displace the extensive ones, accompanied by the emergence of numerous peaks. 
The similar behavior of $n_s$ and $\Omega_m$, and the significant change in filament length distribution for $n_s$, originate from the initial power spectrum, where a larger $n_s$ or $\Omega_m$ corresponds to a smaller amplitude in small k modes.

The sensitivity of filament length distribution to different cosmological parameters persists when we change the number of grid $N_{grid}$ or the smoothing scale $N_{smooth}$ as well as the topological persistence $N_{\sigma}$ in Appendix.\ref{sec:coarse_cosmo}.
 
Notably, the abundance of filaments exhibits a crossover at approximately $60 h^{-1} {\rm Mpc}$ with varying cosmological parameters ($\sigma_8$, $\Omega_m$, $n_s$). This specific value is slightly dependent on different parameter selections, such as $N_{grid}$ and $N_{smooth}$. Therefore, this crossover point signifies a characteristic scale for filaments that warrants further exploration.

The filament length distribution is shown in Fig.~\ref{fig:fit_filament_length}, and it can be well fitted with the following equation Eq.~\ref{eq:filament_fit}, 
\begin{equation}
    log_{10} \Phi_L = -0.002 L^{b} + n_{0}
    \label{eq:filament_fit}
\end{equation}

Here the parameters $b$ and $n_0$ are dependent on the cosmological parameters, and we show the best fitted $b$ and $n_0$ to different cosmological parameters in Fig.~\ref{fig:fit_filament_length_diffcosmo}.
For each cosmology, we use five realization with distinct initial matter distributions to calculate the cosmic variance for $n_0$ and $b$ values and plot 1 $\sigma$ dispersion in the figure. As evident from the left and right panels, the relation is tighter for $\sigma_8$ and $n_s$, while the relation for $\Omega_m$ on the middle panel shows slight stronger sensitivity to cosmic variance.

It is well known from the CDM theory that the number density of massive dark matter haloes is sensitive to cosmological parameters, thus being a powerful probe of cosmology although it is difficult to obtain halo mass from observations. As an interesting theoretical test, here we compare the sensitivity of filament length as a cosmological  probe in comparison to the halo mass function. To determine the halo mass function, we employ the Friends-of-Friends (FOF) method.
Here we use the number density of massive haloes with two mass bins with mass in the range of $\log_{10}(M/M_\odot)\in [13.5,15]$ and $[13.8, 15]$, respectively.
Similarly, for the filament length function, we utilize the fitting parameters $b$ and $n_0$. We define a statistical function as follows:

\begin{equation}
\chi^2=\left(\mathbf{p}_{\text {fiducial }}-\mathbf{p}_{\text {target }}\right) \cdot \mathbf{C o v}^{-1} \cdot\left(\mathbf{p}_{\text {fiducial }}-\mathbf{p}_{\text {target }}\right)
\label{eq:filament_x2}
\end{equation}

Where $\mathbf{p}$ represents the parameter of the statistical quantity, denoting the number density in given mass bin for the halo mass function and the fitting parameters $b$ and $n_0$ for the filament length function. For our analysis, we adopt the Planck 2015 cosmology as the fiducial geometric background. We utilize 40 simulations from the $\Lambda$CDM dataset to compute the covariance matrix and an additional 5 simulations from the $\Lambda$CDM dataset to calculate $\mathbf{p}_{\text{fiducial}}$ for both the halo mass function and the filament length function. Simultaneously, we employ simulations from the $C1$ dataset to calculate $\mathbf{p}_{\text{target}}$.

In Fig.~\ref{fig:x2_filament_length} we compare the ability of constraining cosmological parameters using halo mass function and filament length distribution. The solid lines are for halo mass function with halo mass within [13.5, 15] and the dashed lines are for halo in a higher mass bin of [13.8, 15]. It is found that in general the halo mass function shows stronger constraints on the cosmological parameters of $\sigma_8$ (left panel) and $\Omega_m$ in the right panel. However, the filament length function demonstrates stronger sensitivity to $n_s$ than the halos within the massive bin.  (dashed line in the right panel).

We note that this comparison could just serve as a qualitative consideration. In practice, it is difficult to derive halo mass although it could be done with either x-ray or weak lensing measurements. On the other hand, as mentioned before, one has to use observed galaxies to extract filament. We leave more quantitative comparison to future work.


\subsection{Reshift evolution}
\label{sec:redshift_evolution}

In this section we investigate the evolutionary trend in the distribution of filament lengths and its sensitivity to cosmological parameters. To conduct this analysis, we use 5 redshift snapshots in our numerical simulation: $z=3, 2, 1, 0.5, 0$. Utilizing our filament pipeline, we obtain the filament length distributions for the Planck15 cosmology at each snapshot, with $z=0$ serving as a reference. Fig.\ref{fig:redshift} shows the redshift evolution of the filament length distribution, and it reveals that as redshift decreases from z=3, the number density of filament of all length firstly rises until approximately $z \sim 1$, and from $z \sim 1$ to $z \sim 0$, the number of long filaments declines while short filaments keep increase. To provide a more intuitive understanding, we use $N(L)$ instead of $\Phi_L$, as $N(L)$ less than 1 lack physical meaning. Our findings are in contrast with the previous study by \citet{2014MNRAS.441.2923C}, who reported a predominance of long filaments and fewer short ones at low redshifts. Our results are more consistent with \citet{2024A&A...684A..63G}, despite their use of galaxies as tracers and a different definition of individual filament.
The disagreement with \citet{2014MNRAS.441.2923C} is mainly because they focused only on bifurcation and curvature, neglecting instances where peaks manifest and disrupt the filament, while we use a more broader definition.
Our results reveals a three-step process in filament formation: 1) A first rise in filament formation with time, with rapid formation of longer filament by connections of shorter filaments. 2) After $z\sim 2$, long filaments reach peak efficiency in their formation, leading to a dynamical balance between the disruption and formation of long filament. 3) At lower redshift, increased matter clustering causes some long filaments to fragment, making shorter filaments more prevalent.

Additionally, we also find that Eq.~\ref{eq:filament_beta_fit} can effectively capture the redshift dependence of the filament length function with the following forms,
\begin{equation}
    \begin{aligned}
    log_{10} \Phi_L  &= -0.002 L^{\beta} + n_1\\
    \beta&=a*z^t+b\\
    n_1&=c z^2+d z+e
    \end{aligned}
    \label{eq:filament_beta_fit}
\end{equation}

The relationship between $\beta$ and redshift is illustrated in the left panel of Fig.\ref{fig:beta_redshift}. Specifically, the power-law index within the filament length function demonstrates a power-law correlation with redshift. By employing five redshift values to fit the power-law index and redshift relationship (namely $z=3.0, 2.0, 1.0, 0.5, 0.0$) in black, we successfully predict the behavior for another set of nine redshift values (namely $z=2.2, 1.8, 1.5, 1.2, 0.8, 0.6, 0.3, 0.2, 0.1$) in blue. This suggests that the fitting function effectively describes the power-law index $\beta$ and its relation to redshift, as shown in Fig.\ref{fig:beta_redshift}. Furthermore, we ascertain that the relationship between $n_1$ and redshift can be adequately approximated by a cubic function, which is shown in the right panel of  Fig.~\ref{fig:beta_redshift}. Similarly, it is seen that the parameters obtained from five redshift can well predict the parameters at other redshifts.

Moreover, we observe that an increase in the parameter $\beta$ corresponds to a steep filament length function, indicating a relatively pronounced difference between the numbers of long and short filaments. This suggests a transition from extensive filaments to shorter ones. Furthermore, an increase in the parameter $n_1$ results not only in shorter filaments but also in more extensive ones, thereby increasing the overall presence of filaments. By combining these two relationships, we depict a simple scenario for the formation and evolution of filament in the following.

Under the assumption that long filaments are formed through the merging of shorter ones, we propose a simple picture to describe the evolution of filament. Basically, the formation of short filament is mainly propelled by the growth of density field. Concurrently, short filaments contribute to the maturation of long filaments, whereas localized density enhancements prompt the fragmentation of long filaments into shorter segments. 
Fig.~\ref{fig:filament_model} illustrates our scenario on the formation of both short and long filaments. The (F) process signifies formation of filament arising from the density perturbations, while (B) represents the breakdown of long filament into shorter ones. Additionally, (G) denotes the growth of long filament by merger/connection of short filaments.

Before $z\sim 2$, the $\beta$ value remains below 1, indicating a relatively minor difference between the numbers of long and short filaments. During this period, the dominant processes are filament formation (F) and growth (G), while filament breakdown processes (B) are negligible.
As the redshift approaches $z=2$, the $\beta$ value increases towards 1, signifying that short filaments become more predominant than long filaments. The increasing slope of $n_1$ suggests a significant enhancement in the filament formation process (F). At approximately $z=1.5$, the slope of $n_1$ reaches its peak, indicating the greatest disparity between the growth process (G) and the breakdown process (B), signifying the maximum efficiency of formation of long filaments. Subsequently, the growth process (G) diminishes, while the breakdown process (B) becomes dominant, resulting in the fragmentation of long filaments and the prevalence of shorter ones.

\subsection{Initial condition}
\label{sec:initial}
In the above analysis, we have investigated the impacts of different cosmology parameters on the filament length distribution using simulations in which the initial power spectrum is produced from the CLASS code \citep{2011JCAP...07..034B}. As filament is a distinct structure on large scale, one naive thought is that the number of filaments with a given length could be determined by the power at a given scale. It is then interesting to test how a change of the initial power spectrum at a specific scale will impact the filament length distribution, and inspection of its impact will give more insight of how filament of given length is formed. Here we carry out an experiment by artificially changing the amplitude of the initial power spectrum at different scales with $k \in [0.02, 0.1, 0.2, 0.3,0.4]$. In Fig.\ref{fig:init_power_spectrum} we show an example where we change the amplitude of the initial power spectrum with k in the range $k \in [0.02, 0.1]$. Note that the black line with scale=1 is the standard initial power spectrum, for other lines the power spectrum is increased by a factor given by the parameter scale.

In Fig.~\ref{fig:init_flutuation}, we show the resulting final non-linear power spectrum at z=0 and the filament length distribution for two cases, with the left panels for changing the power on large scale with $k \in [0.02,0.1]$ and the right panels for changing power on small scale with $k \in [0.3,0.4]$. The upper panels show the power spectrum at z=0. It is seen from the left panel that increasing the initial power on large scale, $k \in [0.02,0.1]$, results in strong increase of non-linear power spectrum on small scales. More interesting is that the boost of power spectrum is nearly constant at $k>0.1 {\rm Mpc}^{-1}h$, as can be seen from the lower panel where the deviation of power spectrum respect to the original power spectrum (scale=1) is shown. 
Similar results on the power spectrum are also seen for the right panel with $k \in [0.3, 0.4]$. 
The lower panels of Fig.~\ref{fig:init_flutuation} show the filament length distribution. It is found that increasing the power on large scales leads to formation of more longer filaments. This is because more short filaments are now connected to form long ones, so the total number of filaments, mainly dominated by short filaments, is decreasing. However, the lower right panel shows that increasing the power on small scales has the opposite effect, it leads to formation of more short filaments and less number of long filaments. 

The above results indicate that the effect of power spectrum on the length distribution of filament is complicated. Overall, there is a dynamical equilibrium between the formation of short filament, formation of long filament by connecting short filaments and break of long filament into short ones. Unfortunately, there is lack of theory or model to describe how the length of filament is determined by the power spectrum. Our experimental simulation could serve as a starting point for future development of such a theory or model.

\section{Conclusion}
\label{sect:conclusion}

The large-scale structure of the Universe is in a web like pattern and it contains a wealth of cosmological information. The most prominent structure of the cosmic web is the filamentary structure or the filament, and it has attracted great attention from the community. Different methods have been proposed to find cosmic filament from both the simulation and observational data (see a review by \citet{2018MNRAS.473.1195L}). Existed studies are mainly focusing on either the properties of the filament, such as its length, width and profile (e.g., \citet{2024A&A...684A..63G, 2011MNRAS.414..350S}) or the effects of filament on galaxy properties (e.g., \citet{2011MNRAS.413..971D,2018A&A...609A..84S,2006AJ....131.2332G,2018MNRAS.474.5437L}). Our study in this paper is the first to investigate the effect of cosmological parameters on the distribution of filament, and here we summarize our main results. 

We have developed a filament pipeline, employing COLA for simulating the distribution of dark matter and DisPerSe for filament detection. The parameters are carefully tested to ensure robustness of finding filament. In current study, we construct density field using the distribution of dark matter particles to find filament, so as to avoid any bias from sampling the distribution of galaxies. Undoubtedly, one has to construct density field using observed galaxies for real data. That is to be investigated in future work.

The length distribution of filament at $z=0$ is found to be well described by a power law, with the power index and normalization both depend on cosmological parameters. As dark matter halo mass function is also sensitive to cosmological parameters, we compare the sensitivity of halo mass function and filament length distribution on $\sigma_8$, $\Omega_m$ and the primordial power index $n_s$. It is found that the halo mass function at massive end is more sensitive to $\sigma_8$ and $\Omega_m$, while the filament length is slightly sensitive to $n_s$. 

We also investigate the redshift evolution of filament length. It is uncovered that a three-stage filament formation scenario can well describe the evolution of filament from $z \sim 3$ to $z=0$: rapid formation of both short and long filaments from $z\sim 3$ to $z\sim 2$, and a continuous increasing of short filaments and a stable persistence of long filament between $z\sim 2$ to $z\sim 1$. Below $z\sim 1$, the number of short filaments is mild increasing, but the number of long filament is decreasing by fragmentation/breaking into short filaments. The filament length evolution can still be well described by the power law, with both the power index and normalization are described using Eq.\ref{eq:filament_beta_fit}.

Finally, we test how the shape of the initial power spectrum affects the filament length distribution. By artificially modifying the power of the initial perturbation at a given scale, it is found that increasing the initial power spectrum on large scale leads to increasing of long filaments and decrease of short filaments. Increase of initial power on small scales leads to more formation of short filaments and decrease of long filament. Thus, it is not only the normalization of the initial power spectrum, but its shape both affect the formation of filament.

In summary, our study demonstrates that the length of filament could serve as a direct and effective cosmological probe. However, as we noted before, our study is based on the distribution of dark matter, while in reality one has to use the distribution of galaxies, and it calls for further study on the use of real data for cosmological constraints. On the other hand, as we have seen, there is lack of model or theory to describe the length distribution of filament and its cosmological dependence, unlike the classic theory for the halo mass function \citep{1974ApJ...187..425P,1996Natur.380..603B}. It is of great importance to explore a theoretical modelling of filament length distribution in future, and our experimental simulation could be seen as a first step towards this goal.

\section*{Acknowledgements}
We thank the anonymous referee for constructive comments to improve the quality of our paper. We also acknowledge the support from the National Key Research and Development Program of China (No.2022YFA1602903), the NSFC \ (No. 11825303, 12347103, 11861131006), the science research grants from the China Manned Space project with No. CMS-CSST-2021-A03, CMS-CSST-2021-B01, the Fundamental Research Funds for the Central Universities of China \ (226-2022-00216) and the start-up funding of Zhejiang University. RC acknowledges in part financial support from the start-up funding of Zhejiang University and Zhejiang provincial top level research support program. We also acknowledge the cosmology simulation database (CSD) in the National Basic Science Data Center (NBSDC) and its funds the NBSDC-DB-10.

\section*{Data Availability}
The simulation datasets generated in the current study are available from the corresponding authors on reasonable request.




\bibliographystyle{mnras}
\bibliography{example} 




\appendix
\section{the choosing of $N_{\sigma}$}
\label{sec:nsigma}

The DisPerSE method incorporates an $N_{\sigma}$ parameter that characterizes topological persistence, aiding in the identification of robust components of the filamentary network in comparison to a discrete random Poisson distribution. A higher $N_{\sigma}$ value facilitates the exclusion of less significant critical points which are considered as noise, thereby preserving only the most topologically robust features. However, a higher value of $N_{\sigma}$ also reduces the completeness of filament detection, as some less significant filaments may be mistakenly categorized as noise.

\begin{figure}
	\includegraphics[width=\columnwidth]{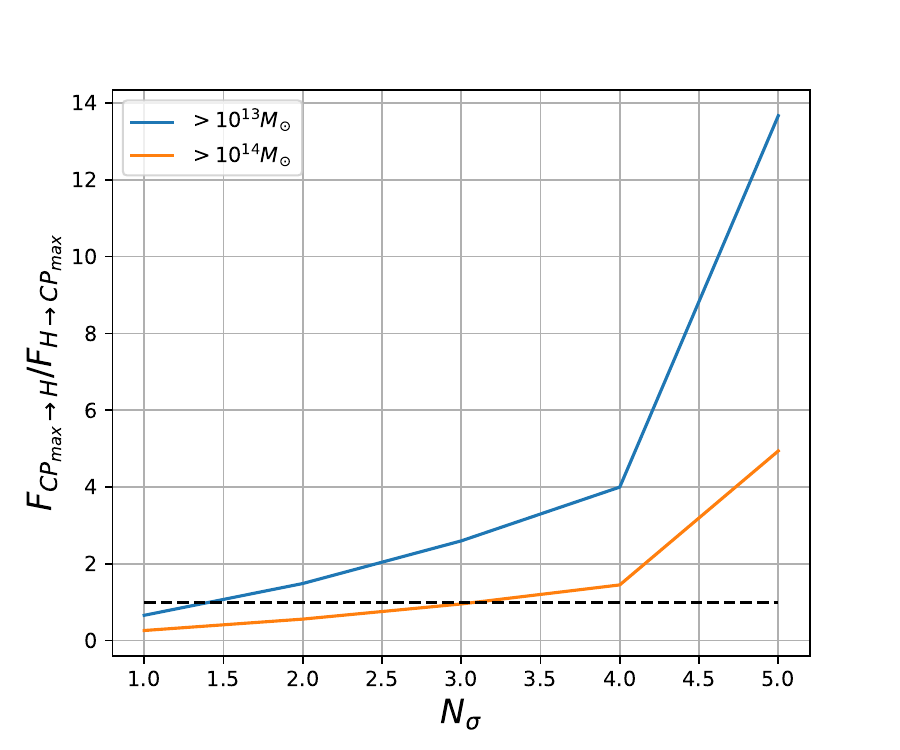}
   \caption{Assessment of the robustness of the variable parameter $N_{\sigma}$ in the filament pipeline following the methodology of \citet{2024arXiv240204837B}.} 
    \label{fig:cph}
\end{figure}

The $CP_{\text{max}}$-Halo calibration method can be utilized for calibration. This method involves comparing the positions of nodes ($CP_{\text{max}}$) to those of the most massive halos identified by the Friend-of-Friends (FoF) algorithm, as depicted in Fig.~\ref{fig:cph}. Following \citet{2024arXiv240204837B}, we present the ratio of the fraction of $CP_{\text{max}}$ nodes located within the virial radius ($r_{\text{vir}}$) of any massive halo to the fraction of massive halos hosting a $CP_{\text{max}}$ node within their $r_{\text{vir}}$ spheres. This comparison aims to identify the topological persistence value that brings the ratio closest to unity, minimizing both the number of halos without node associations and the number of nodes without halo associations at $z = 0$. Consequently, we select $N_{\sigma} = 3$. Notably, we define the most massive halos as those with masses exceeding $10^{14} M_{\odot}h^{-1}$ to ensure compatibility with the smoothing scale encompassing $10^{13} M_{\odot}h^{-1}$.

\begin{figure}
	\includegraphics[width=\columnwidth]{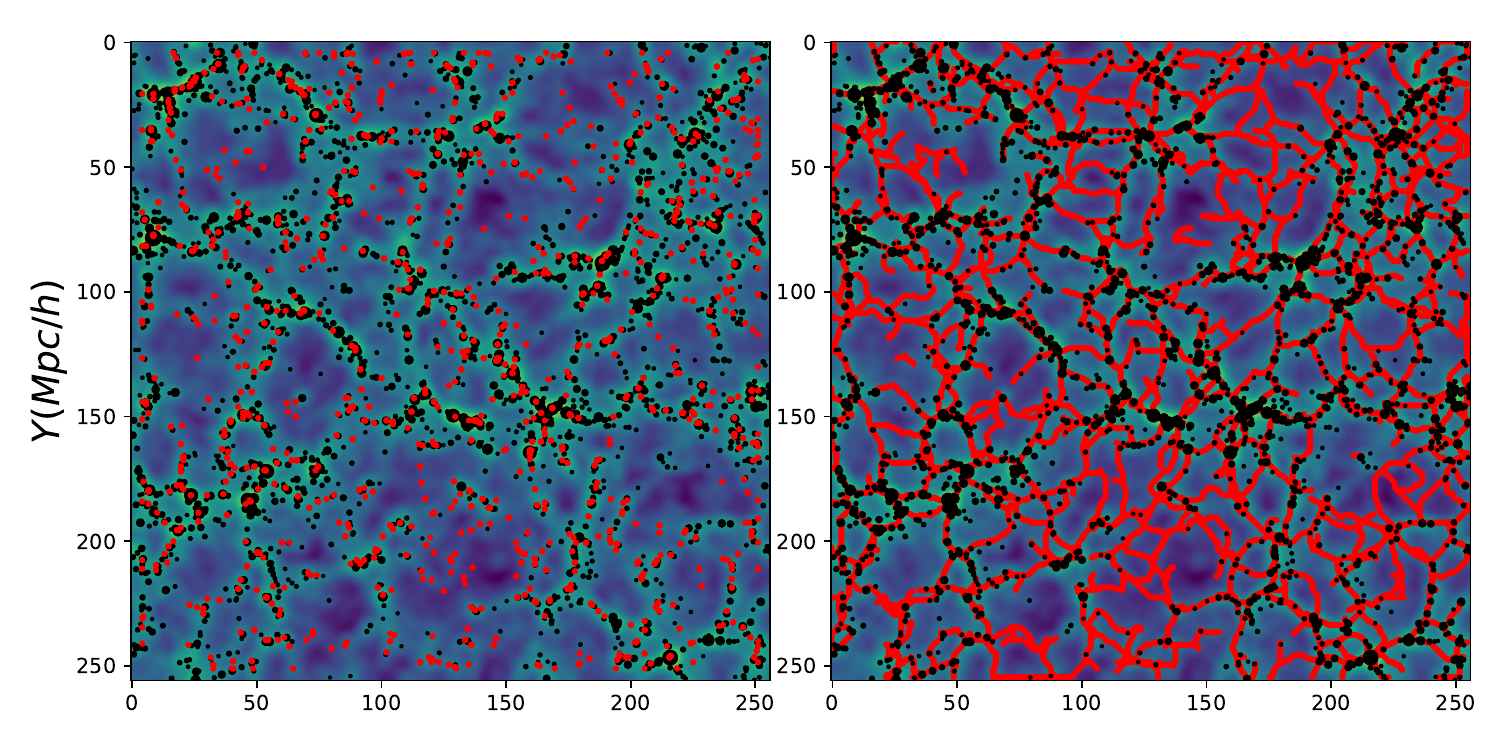}\\
        \includegraphics[width=\columnwidth]{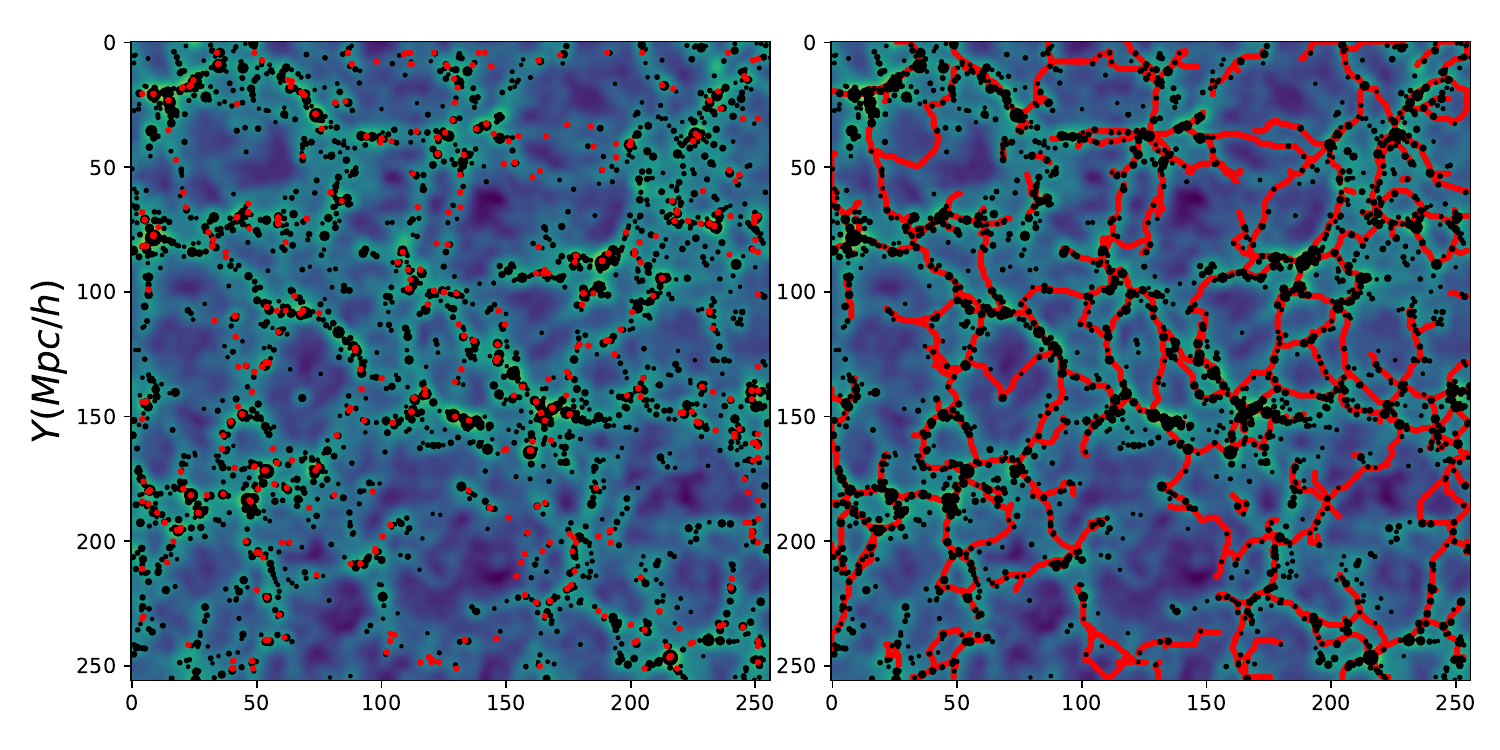}\\
        \includegraphics[width=\columnwidth]{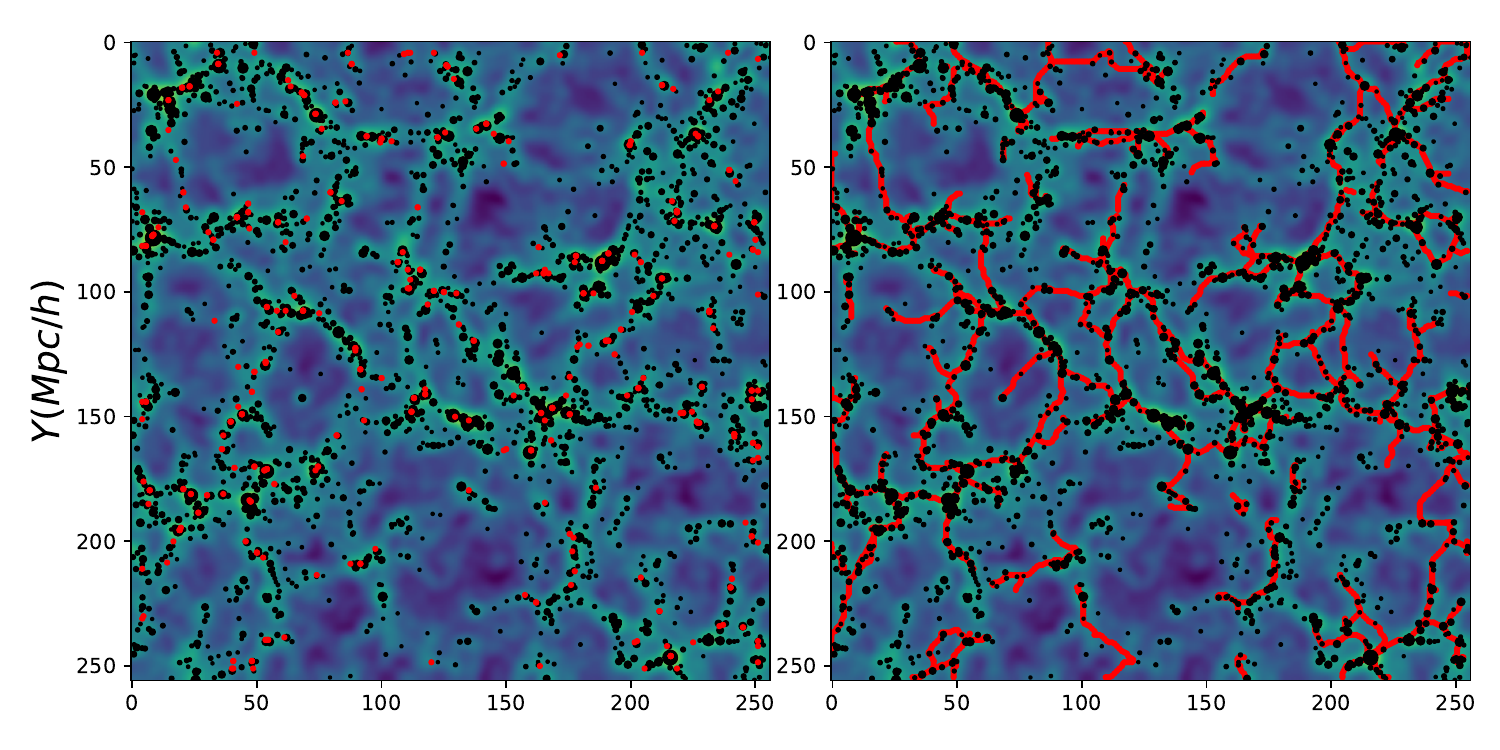}\\
        \includegraphics[width=\columnwidth]{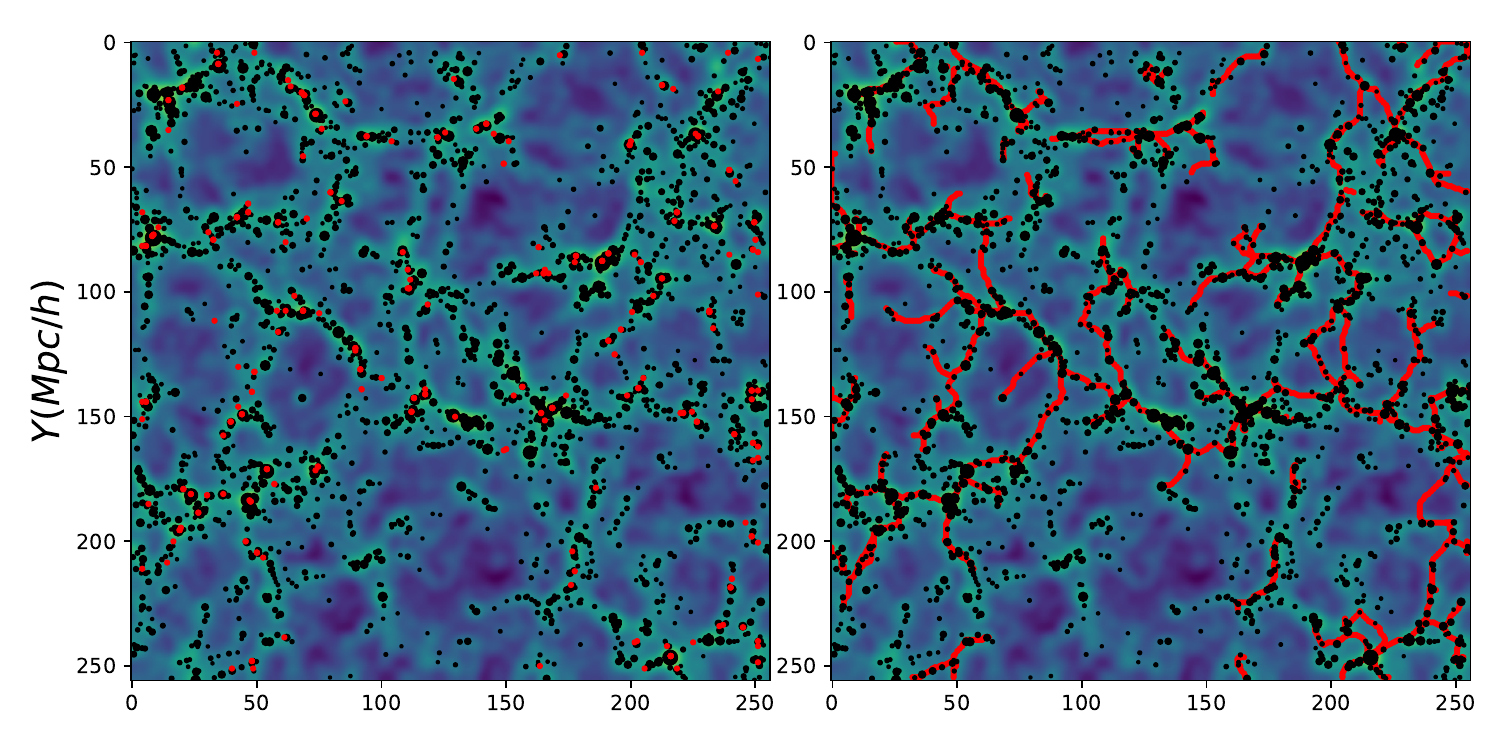}
        \includegraphics[width=\columnwidth]{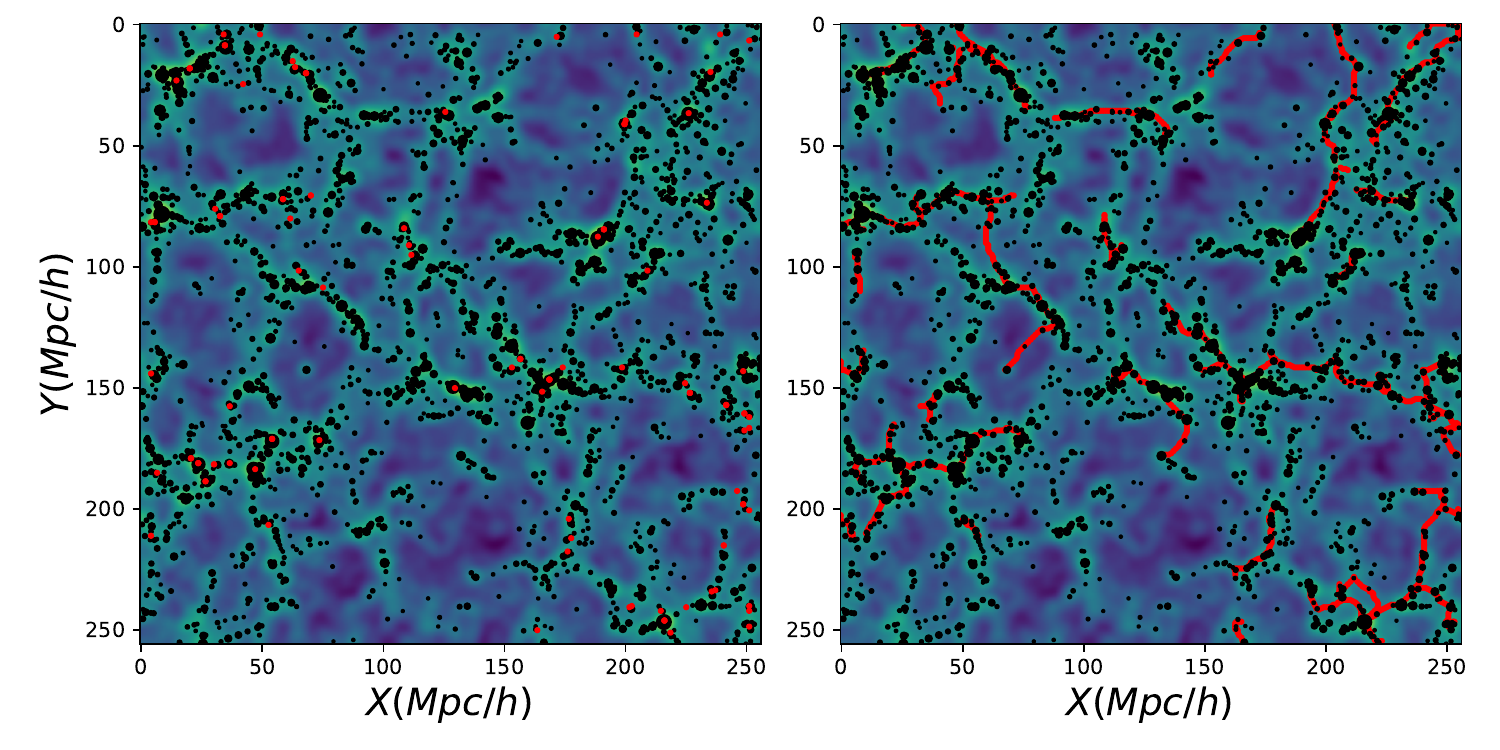}
   \caption{Visual comparison of different persistence parameters $N_{\sigma}$. A slice of the density field with size of $20*256*256 (h^{-1} {\rm Mpc})^3$ with $N_{\sigma} = 1,2,3,4,5$ from top to bottom.} 
    \label{fig:sigma_graph}
\end{figure}

\begin{figure}
	\includegraphics[width=\columnwidth]{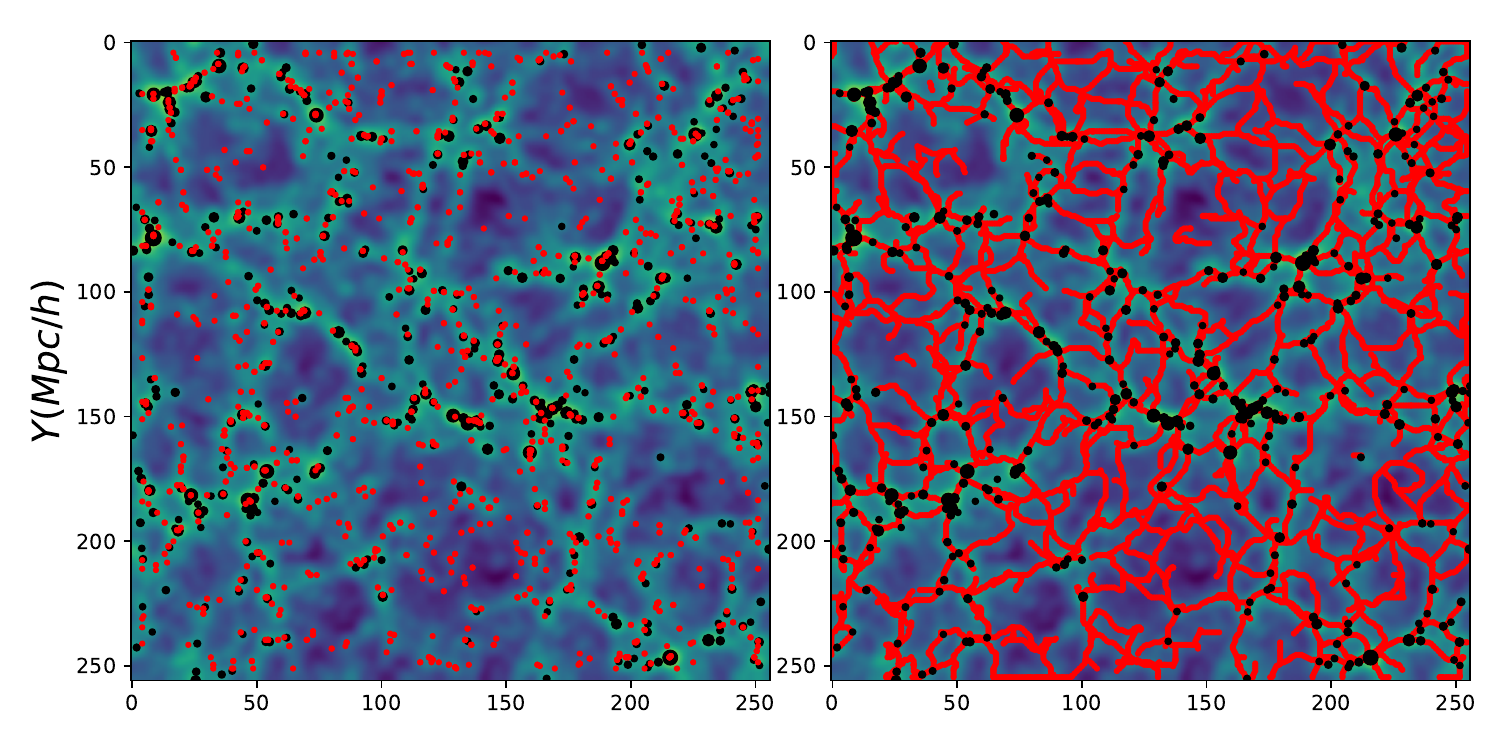}\\
        \includegraphics[width=\columnwidth]{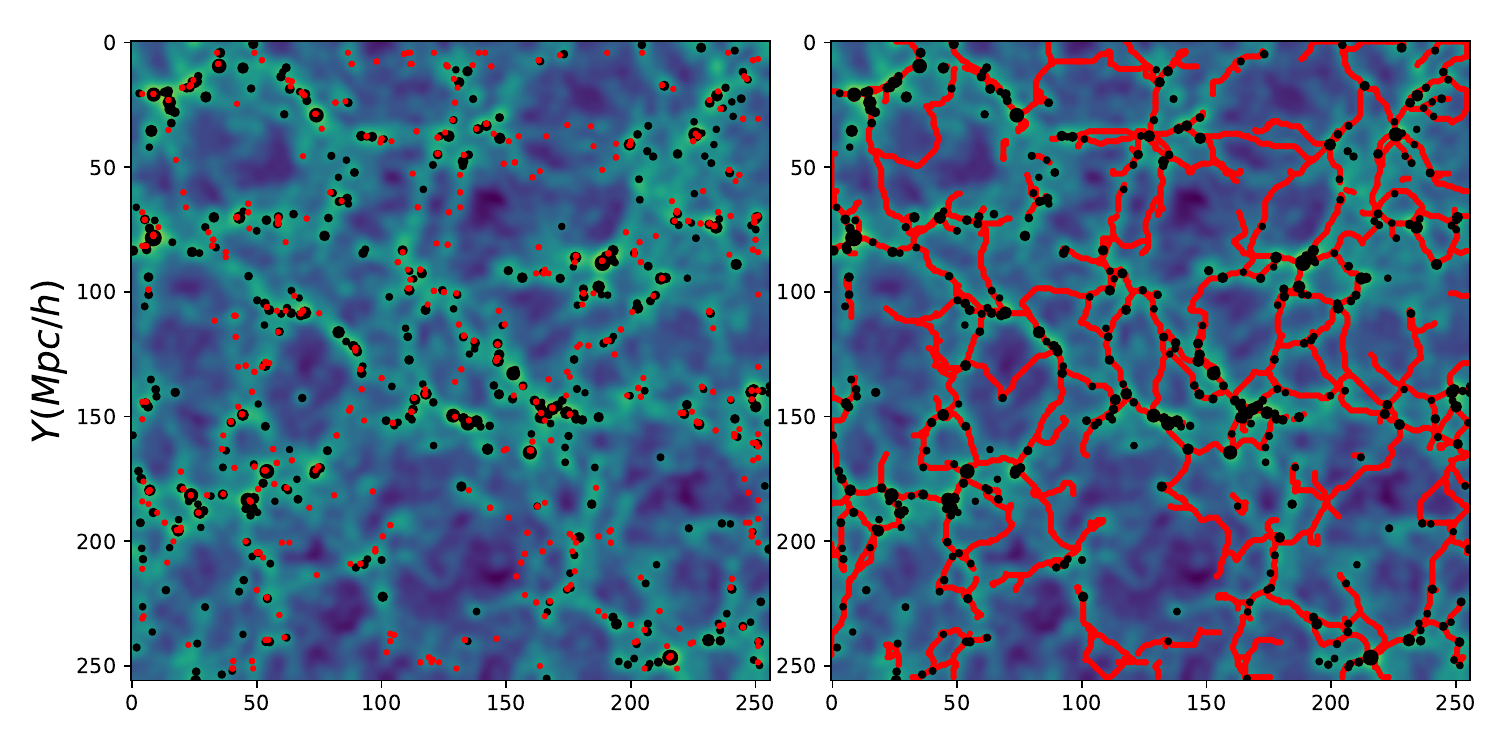}\\
        \includegraphics[width=\columnwidth]{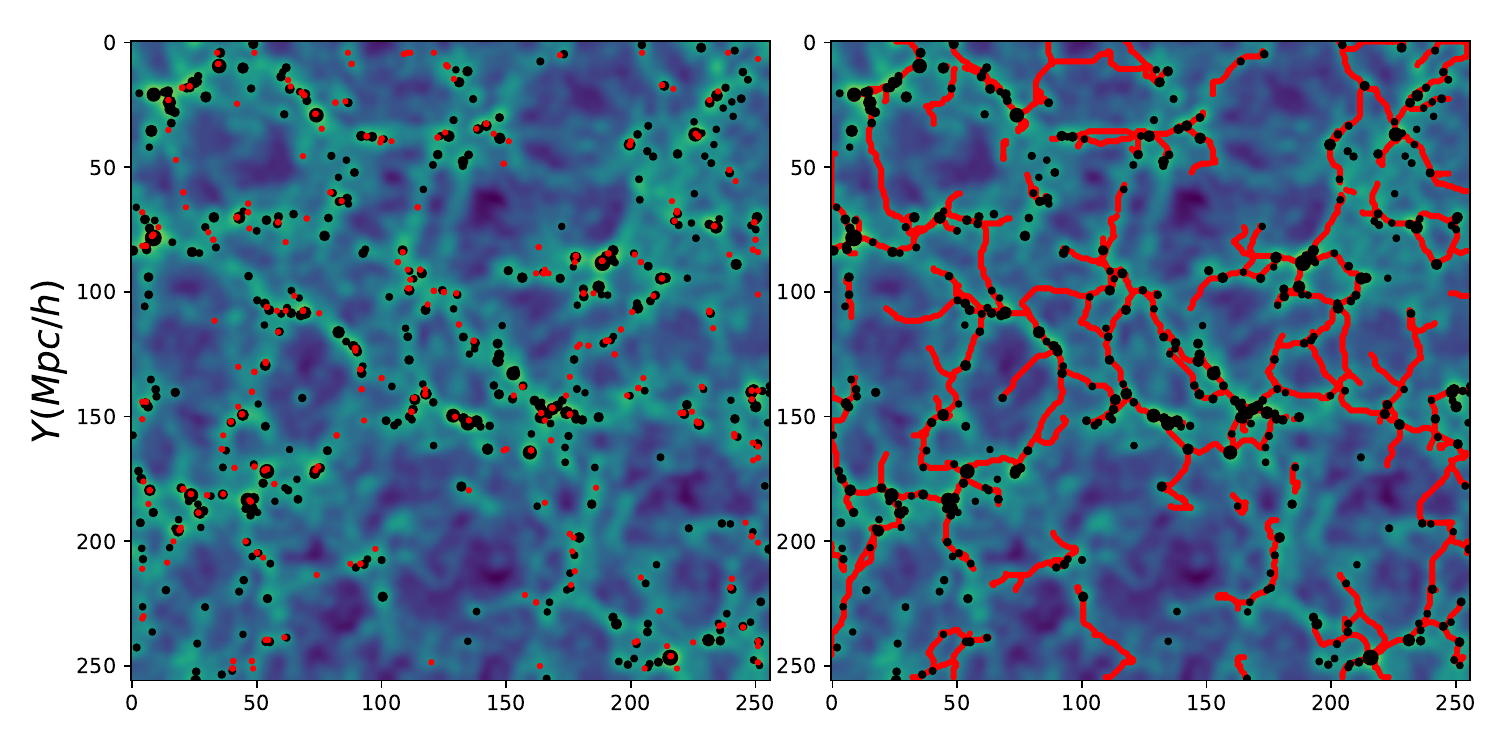}\\
        \includegraphics[width=\columnwidth]{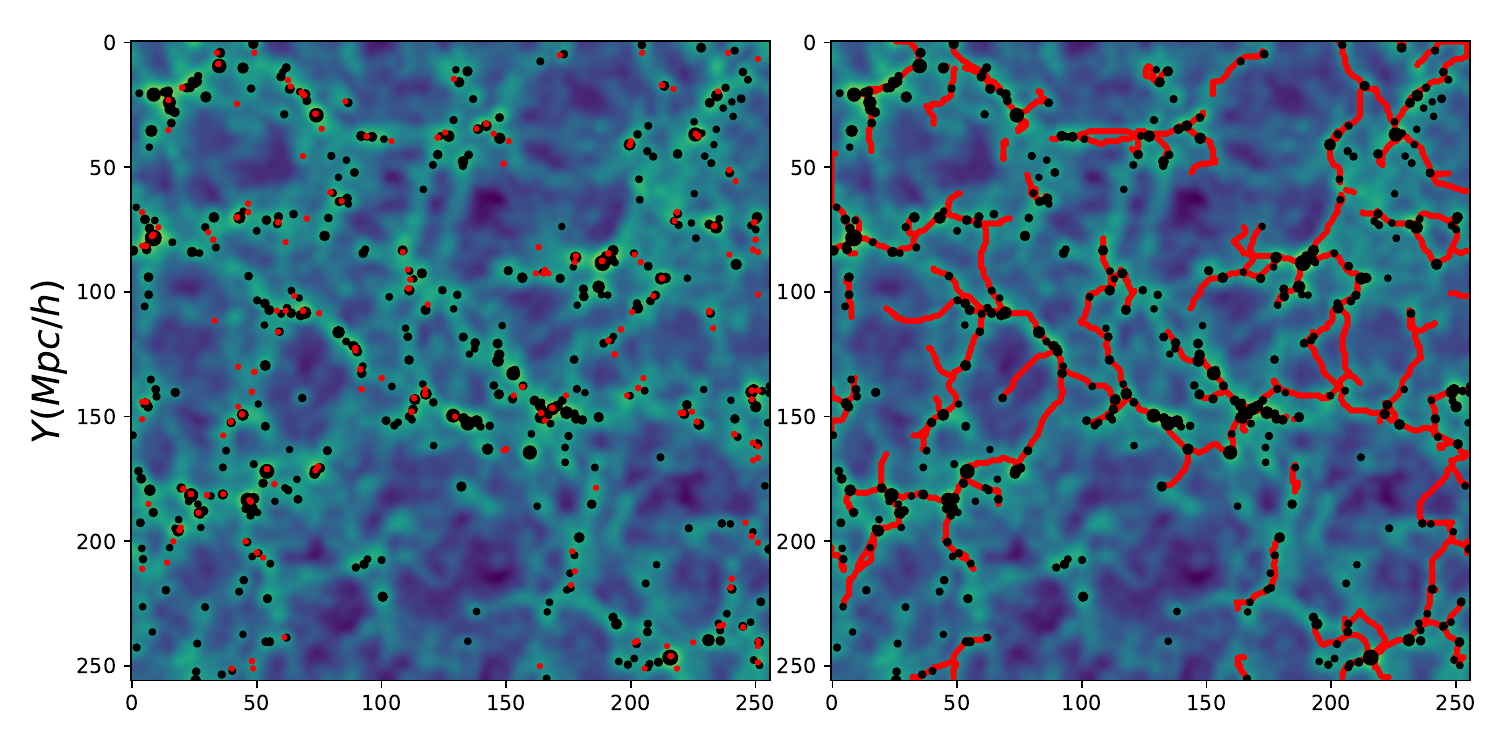}
        \includegraphics[width=\columnwidth]{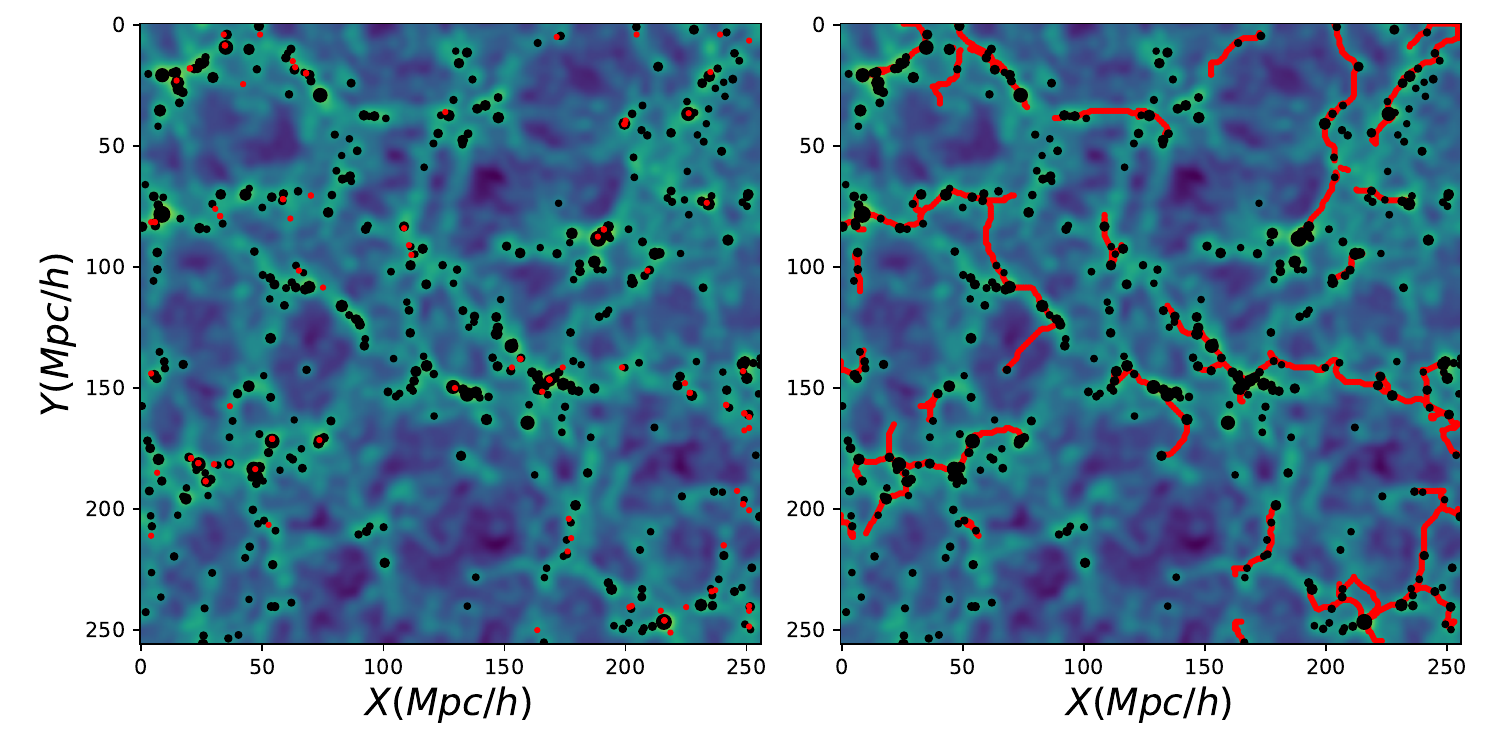}
   \caption{Similar to Fig. \ref{fig:sigma_graph}, but only the most massive Friend-of-Friends (FoF) halos (with masses $> 10^{14} M_{\odot}h^{-1}$) are represented by black dots.} 
    \label{fig:sigma_graph2}
\end{figure}

\begin{figure*}
\centering
	\includegraphics[width=\textwidth]{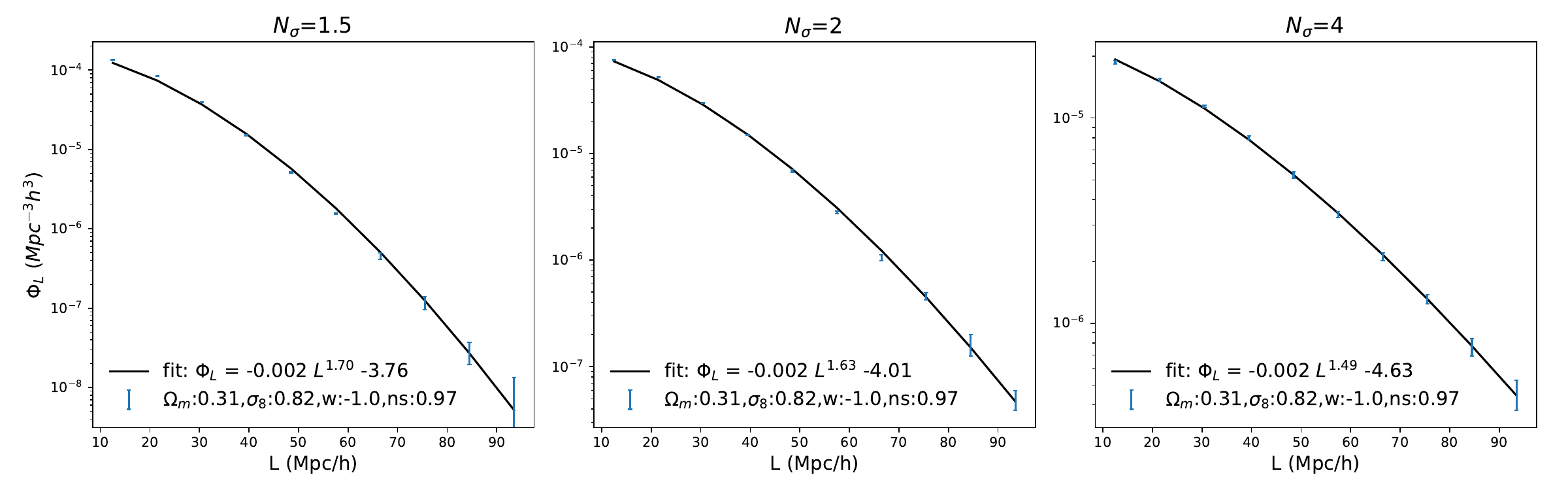}
   \caption{Our fitting model of the filament length function for different choice of parameters $N_\sigma$, depicted as $1.5, 2, 4$ from left to right. The figure shows the robustness of our fitting model across various $N_\sigma$.}
    \label{fig:fit_s15}
\end{figure*}

However we have also determined that the $N_{\sigma}$ value is sensitive to the mass threshold of massive halos. Altering the threshold to $10^{13} M_{\odot}h^{-1}$ necessitates a corresponding adjustment of $N_{\sigma}$ to 1.5 to maintain the unity ratio. Nevertheless, further exploration of the calibration method is warranted. To validate our choice, we utilize visual comparison methods as described in \citet{2020A&A...641A.173G}. 
Fig.~\ref{fig:sigma_graph} shows slices at \(z=0\) for various \(N_{\sigma}\) values, arranged from top to bottom. In the left panel, \(CP_{\text{max}}\) nodes are depicted as red dots, while Friend-of-Friends (FoF) halos (with masses > \(10^{13} M_{\odot}h^{-1}\)) are represented by black dots. Similarly, the right panel displays filaments as red lines, with halos indicated by black dots; larger dots correspond to higher halo masses. In contrast, Fig.~\ref{fig:sigma_graph2} focuses only on the most massive FoF halos (masses > \(10^{14} M_{\odot}h^{-1}\)), represented by black dots. Notably, selecting \(N_{\sigma} = 1\) or \(N_{\sigma} = 2\) leads to the detection of numerous $CP_{\text{max}}$ nodes in non-massive regions, which is undesired. Thus, we affirm that $N_{\sigma} = 3$ is a suitable choice for our study.
Furthermore, we confirm the robustness of our filament length function fitting model across different $N_{\sigma}$ selections, as shown in Fig.~\ref{fig:fit_s15}. We chose three different values of $N_{\sigma}$ (1.5, 2, and 4), all of which fit well, indicating that our conclusions are not sensitive to this parameter.

\section{The cosmological sensitivity for different parameter $N_{grid}$, $N_{smooth}$ and $N_{\sigma}$}
\label{sec:coarse_cosmo}

\begin{figure*}
\centering
	\includegraphics[width=0.32\textwidth]{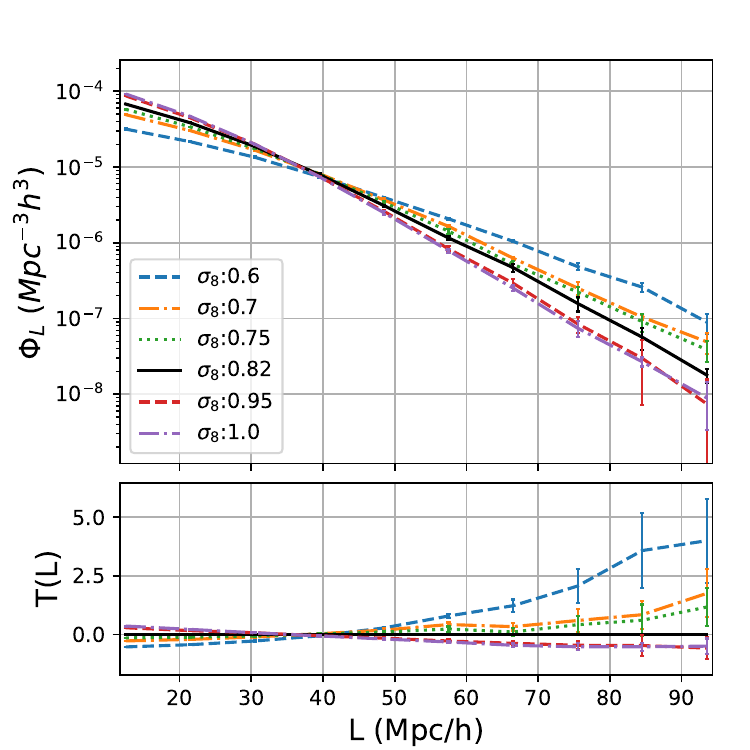}
        \includegraphics[width=0.32\textwidth]{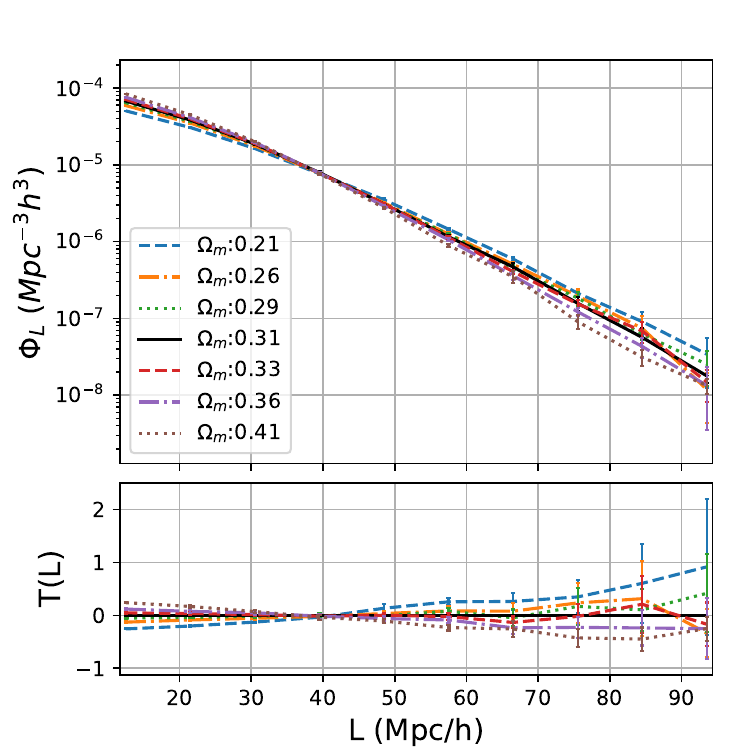}
        \includegraphics[width=0.32\textwidth]{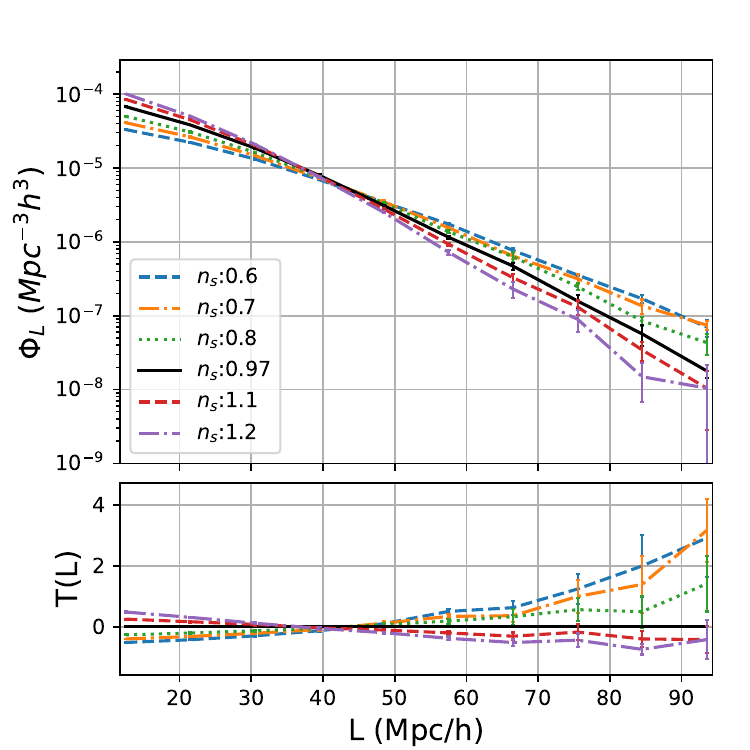}\\
        \includegraphics[width=0.32\textwidth]{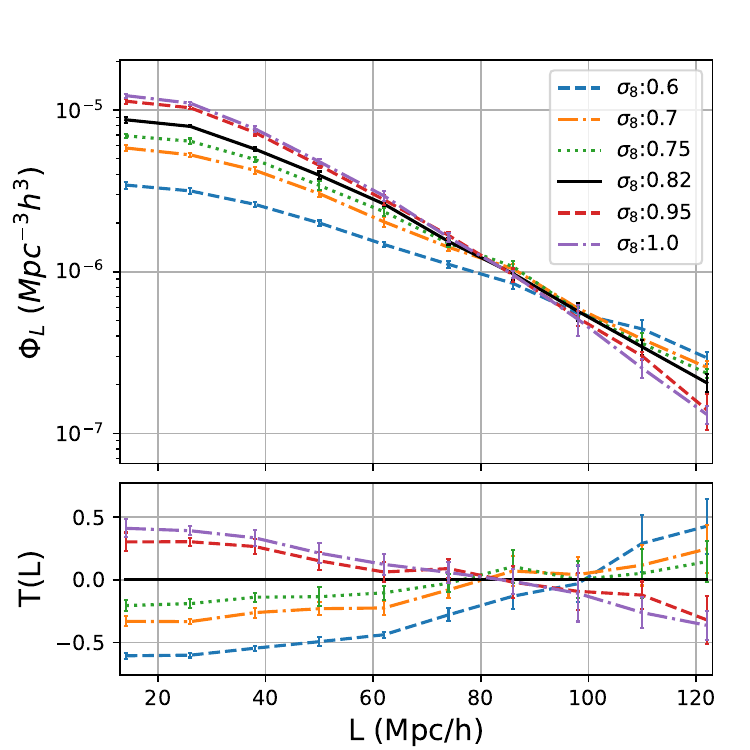}
        \includegraphics[width=0.32\textwidth]{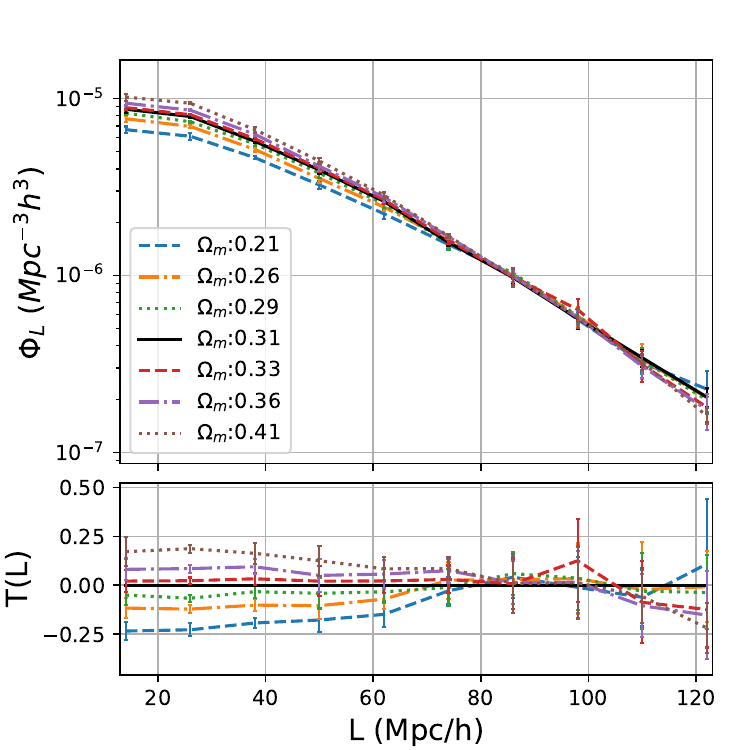}
        \includegraphics[width=0.32\textwidth]{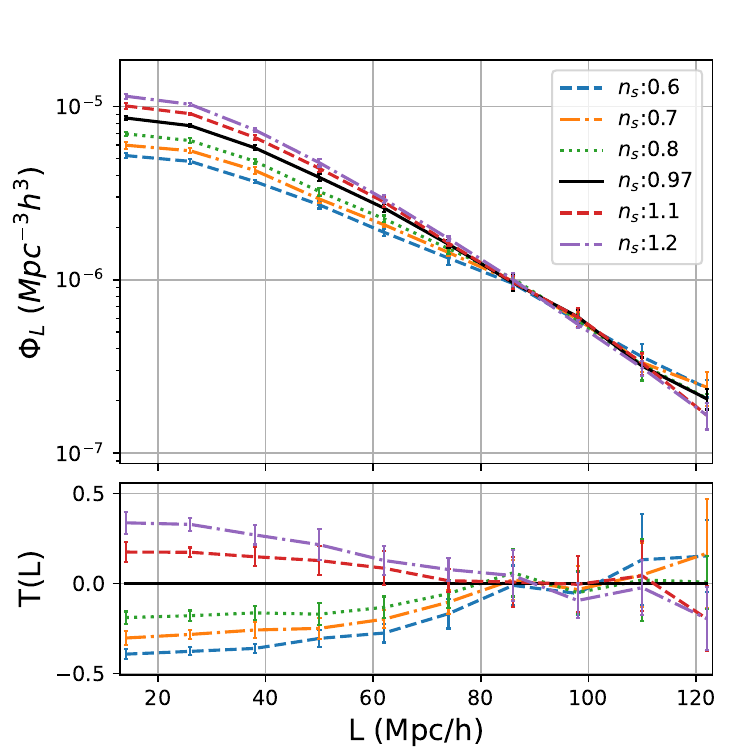}
   \caption{Filament length distribution computed from the filament pipeline for different cosmologies, using either a small lattice number $N_{grid}$ or a large smooth scale $N_{smooth}$. See text for more details.}
    \label{fig:coarse_cosmology}
\end{figure*}

\begin{figure*}
\centering
	\includegraphics[width=0.32\textwidth]{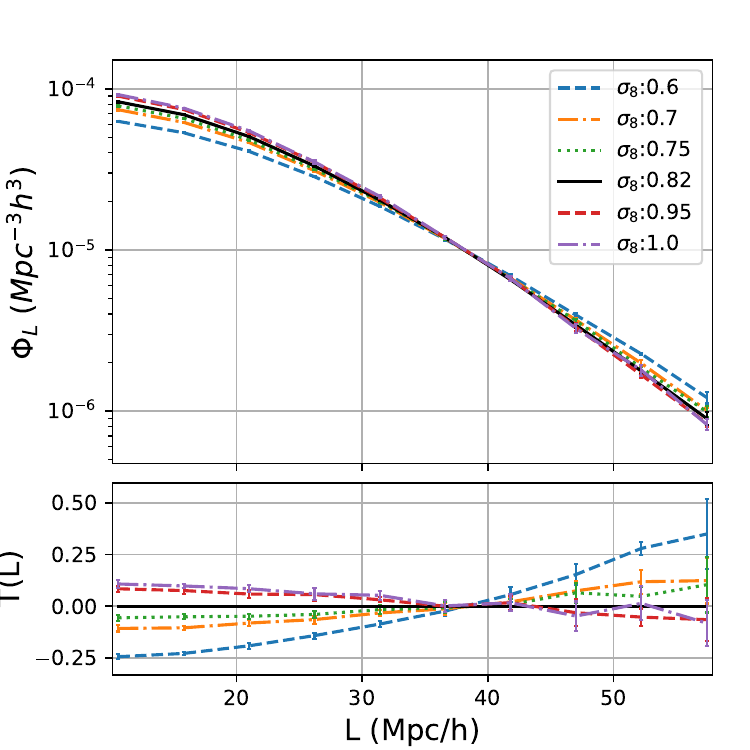}
        \includegraphics[width=0.32\textwidth]{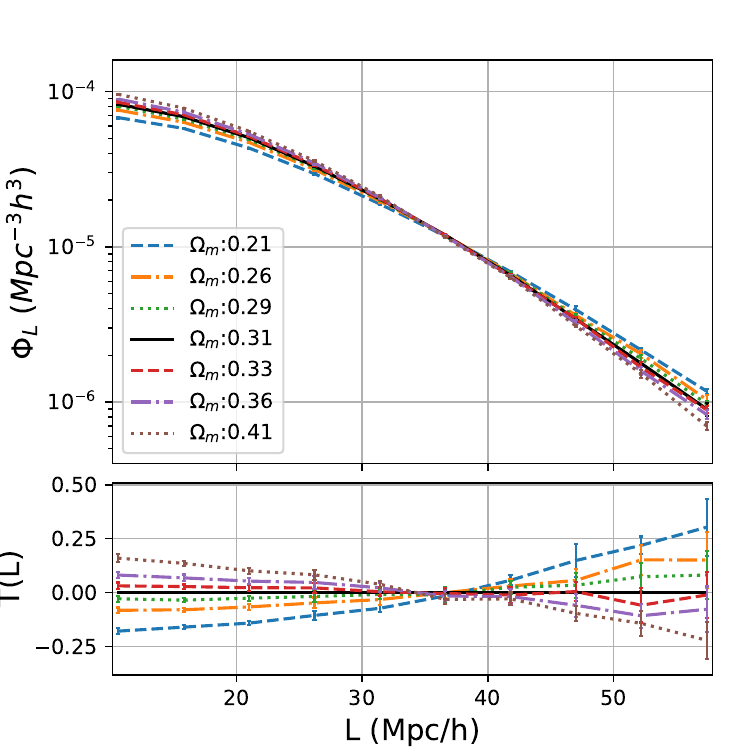}
        \includegraphics[width=0.32\textwidth]{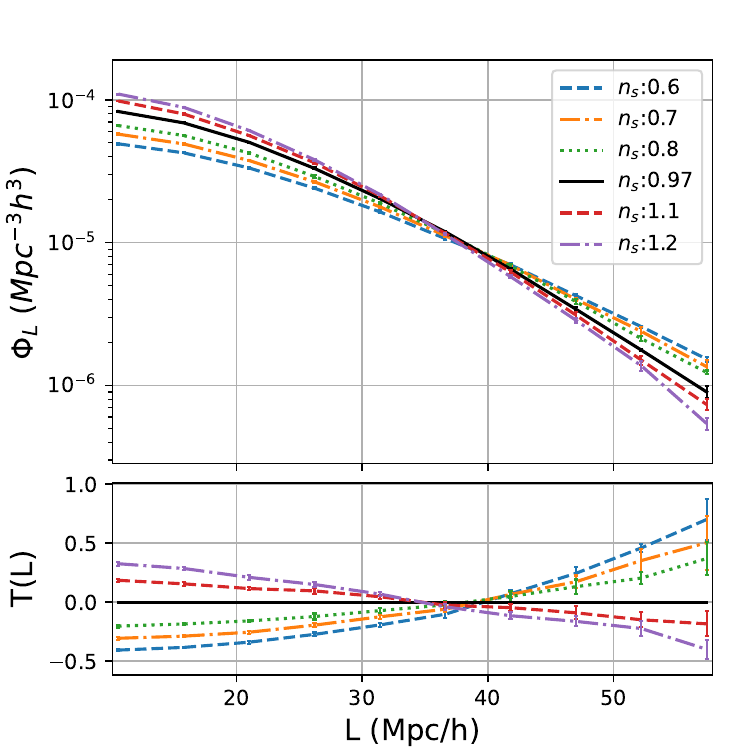}
    \caption{Filament length distribution computed from the filament pipeline for different osmological parameters, utilizing the persistence parameter $N_\sigma = 1.5$. The left, middle, and right panels display the filament length functions for varying $\sigma_8$, $\Omega_m$, and $n_s$, respectively.}
    \label{fig:cosmo_s15}
\end{figure*}

We examine the cosmological dependence of the filament length function with various parameters: the number of grid points $N_{\text{grid}}$, the smoothing scale of the field $N_{\text{smooth}}$, and the topological persistence $N_{\sigma}$. In Fig.~\ref{fig:coarse_cosmology}, we compute the filament length function using different cosmological parameters, varying $\sigma_8$, $\Omega_m$, and $n_s$ from left to right. The top panel shows results using $N_{\text{grid}} = 128$, indicating lower resolution compared to our default setting $N_{\text{grid}} = 512$, yet the cosmological dependence on filament length remains consistent. Similar results are observed when varying the smoothing scale $N_{\text{smooth}}$ from $2h^{-1}\rm{Mpc}$ to $3h^{-1}\rm{Mpc}$ in the bottom panel of Fig.~\ref{fig:coarse_cosmology}, confirming persistent cosmological dependencies across different density field constructions. In Fig.~\ref{fig:cosmo_s15}, adjusting the topological persistence $N_{\sigma}$ from 3 to 1.5 alters the most massive halos from $10^{14}M_{\odot}h^{-1}$ to $10^{13}M_{\odot}h^{-1}$ in the $CP_{\rm{max}}$-$Halo$ calibration method, yet the observed cosmological dependence persists, indicating robust findings across varying parameters in DisPerSE. Hence, the cosmological dependence pattern remains robust across different parameter settings.

\bsp	
\label{lastpage}
\end{document}